\documentclass[twoside]{article}

\usepackage[accepted]{aistats2025}

\RequirePackage{algorithm}
\RequirePackage{algorithmic}
\usepackage{graphicx}
\usepackage{amsfonts} 
\usepackage{dsfont} 
\usepackage{amsthm}
\usepackage{thmtools}
\usepackage{thm-restate}
\usepackage{hyperref}
\usepackage{cleveref}
\usepackage{bm}
\usepackage{booktabs}
\usepackage{makecell}
\usepackage{stmaryrd}
\usepackage{mathtools}
\newtheorem{theorem}{Theorem}[section]

\newtheorem{corollary}[theorem]{Corollary}
\newtheorem{remark}[theorem]{Remark}

\usepackage{geometry}
\DeclareMathOperator*{\argmin}{arg\,min}

\newcommand{\inv}{^{\raisebox{.2ex}{$\text{-}\scriptscriptstyle\!1$}}}
\newcommand\numberthis{\addtocounter{equation}{1}\tag{\theequation}}
\usepackage{url}

\usepackage[round]{natbib}

\begin{document}

%
\runningtitle{Cross-modality Matching and Prediction of Perturbation with Labeled Gromov-Wasserstein OT}

%
\runningauthor{Ryu, Bunne, Pinello, Regev, and Lopez}

\twocolumn[

\aistatstitle{Cross-modality Matching and Prediction of Perturbation Responses with Labeled Gromov-Wasserstein Optimal Transport}

\aistatsauthor{ Jayoung Ryu \\ Genentech\\Harvard Medical School\\Massachusetts General Hospital \And Charlotte Bunne \\ Genentech\\Stanford University \And Luca Pinello \\  Massachusetts General Hospital\\Broad Institute\\Harvard Medical School \AND Aviv Regev$^*$ \\Genentech  \And Romain Lopez$^{*}$\\ Genentech\\Stanford University}
\vspace{1cm}
]



\begin{abstract}
It is now possible to conduct large scale perturbation screens with complex readout modalities, such as different molecular profiles or high content cell images. While these open the way for systematic dissection of causal cell circuits, integrating such data across screens to maximize our ability to predict circuits poses substantial computational challenges, which have not been addressed. Here, we extend two Gromov-Wasserstein optimal transport methods to incorporate the perturbation label for cross-modality alignment. The obtained alignment is then employed to train a predictive model that estimates cellular responses to perturbations observed with only one measurement modality. We validate our method for the tasks of cross-modality alignment and cross-modality prediction in a recent multi-modal single-cell perturbation dataset. Our approach opens the way to unified causal models of cell biology. 
\end{abstract}

\section{INTRODUCTION}

High-throughput high-content perturbation screens are revolutionizing our ability to interrogate gene function and identify the targets of small molecules~\citep{bock_high-content_2022}. Advances over the past decade now allow us to efficiently measure the responses of individual cells to tens of thousands of perturbations in terms of different molecular profiles, such as RNA~\citep{Dixit2016-kw}, chromatin and/or proteins~\citep{frangieh2021multimodal} (Perturb-Seq) or high content microscopy images~\citep{Feldman2019-vb} (Optical Pooled Screens (OPS)). Because these modalities provide complementary information about how cells respond to perturbation, leveraging multi-modal data with adequate computational methods presents a remarkable opportunity to learn of all aspects of cell biology. First, relating rich molecular profiles to cell biology morphological phenotypes helps understand how different levels of organization relate to each other in the cell, a fundamental question in biology. Second, cross-modality translation methods could help reduce experimental costs and speed up discovery. In particular, molecular profiling (Perturb-Seq) screens are usually substantially more costly and less scalable than optical pooled screens, but provide far more mechanistically interpretable results.  Predicting molecular profiles (e.g., RNA profiles) from morphological profiling data would thus offer both scalability and interpretability, accelerating discovery. 

Cross-modality alignment and prediction are well-studied tasks in (non-perturbational) single-cell genomics. Multiple approaches based on autoencoders \citep{Ashuach2023-xu} and Gromov-Wasserstein Optimal Transport (GWOT) \citep{Demetci2022-qi,Demetci2022-xe} have successfully tackled these problems for datasets where cell states are clearly demarcated (e.g., discrete cell types). However, data from single-cell perturbation studies are significantly more challenging, because screens are usually conducted in one, relatively homogeneous, cell type, and each perturbation typically only induces a relatively modest change in the cell's overall state. In such cases, existing alignment methods may perform poorly. 

\begin{figure*}[ht]
\vspace{-.1in}
\begin{center}
\centerline{\includegraphics[width=1.4\columnwidth]{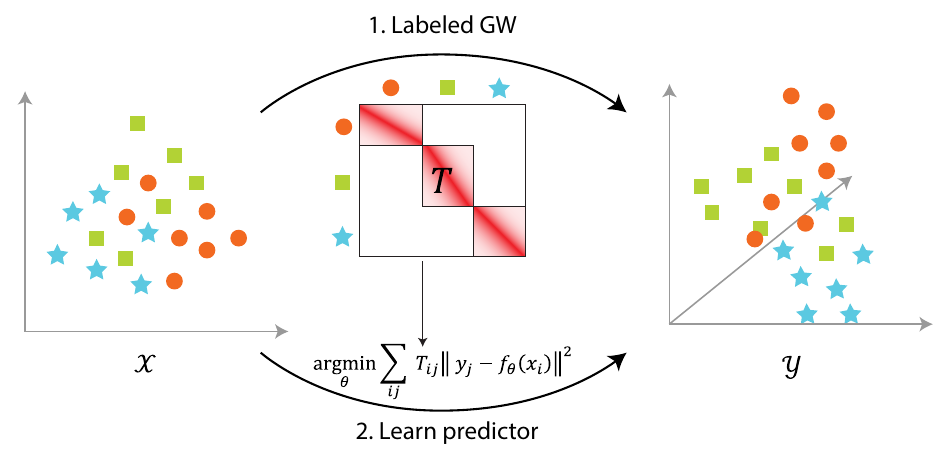}}
\vspace{-.1in}
\caption{Schematic of the proposed computational workflow.}
\label{schematic}
\end{center}
\vspace{-.3in}
\end{figure*}

To tackle this, we propose to leverage the perturbation label for each cell, readily available for such data, to infer a more accurate cross-modality alignment. The naive approach of aligning cells across modalities for each perturbation separately is sub-optimal, because multiple different perturbations can cause similar phenotypic shifts \citep{Dixit2016-kw,frangieh2021multimodal,Geiger-Schuller2023.01.23.525198}, such that samples from other perturbations should provide information about the global topology of the phenotypic space. Instead, we incorporate the label information when learning a model across all the perturbations, and show this substantially improves the performance. Specifically, we adapt GWOT methods, including entropic GWOT ~\citep{Peyre2016-jg} and Co-Optimal Transport (COOT),~\citep{Redko2020-bn} to exploit this information as a constraint on the learned cross-modality cellular coupling matrix. We then employ the learned coupling matrix to train a cross-modality prediction model (Figure \ref{schematic}) and apply it to estimate the response to perturbations observed only in one modality (out-of-sample). We validate our method by benchmarking against baselines in recent data from a multi-modal small molecule screen.

\section{RELATED WORK}

We first introduce Gromov-Wasserstein Optimal Transport as an important technical foundation for multi-modal data alignment. We then discuss extensions that incorporate additional structural information, as well as applications to domain adaptation problems.

\subsection{Gromov-Wasserstein Optimal Transport}

Let us consider two discrete measures $\mu=\sum_{i=1}^n{p_i \delta_{x_i}}$, and $\nu=\sum_{j=1}^m{q_j \delta_{y_j}}$ with supports $\mathcal{X}$ and $\mathcal{Y}$, respectively. Let $C \in \mathbb{R}^{n \times m}$ be the cost matrix where $C_{ij} = c(x_i, y_j)$ for the cost function $c: \mathcal{X} \times \mathcal{Y} \rightarrow \mathbb{R}_+$. The Optimal Transport (OT) problem consists of finding the transport plan (or \textit{coupling}) $T^* \in \mathcal{C}_{p,q}$, where:
\begin{align}
\mathcal{C}_{p,q} = \{T \in \mathbb{R}_+^{n \times m} \mid T\mathbf{1}_m = p, T^\top\mathbf{1}_n = q\},
\end{align}
that minimizes the cost of transporting $\mu$ onto $\nu$ ~\citep{Monge1781-md, Kantorovich1960-fe}:
\begin{restatable}{eq}{ot}
    \begin{equation}
        \mathcal{OT}(\mu, \nu) = \min_{T \in \mathcal{C}_{p,q}} \langle C, T \rangle \label{eq:OT}\tag{OT}
    \end{equation}
\end{restatable}
where $\langle C, T \rangle = \sum_{i,j} C_{ij}T_{ij}$ denotes the Frobenius inner product between matrices $C$ and $T$.
The cost function is easily defined when both $\mathcal{X}$ and $\mathcal{Y}$ are subsets of a normed vector space (e.g., $c$ is the distance induced by the norm).

For multi-modal alignment, however, $\mathcal{X}$ and $\mathcal{Y}$ belong to incomparable spaces. In this scenario, it is more relevant to consider the Gromov-Wasserstein Optimal Transport (GWOT) distance $\mathcal{GW}(\mu, \nu)$, which employs as its cost function the \textit{distance between distances}~\citep{Memoli2011-fq, Alvarez-Melis2018-zx}:
\begin{align}
\min_{T\in\mathcal{C}_{p,q}}{\sum_{i,j, k,l} d\left(c_{\mathcal{X}}(x_i, x_k),c_{\mathcal{Y}}(y_j, y_l)\right) T_{ij} T_{kl}},
\end{align}
with within-domain cost functions $c_\mathcal{X}: \mathcal{X} \times \mathcal{X} \rightarrow \mathbb{R}_+$, $c_\mathcal{Y}: \mathcal{Y} \times \mathcal{Y} \rightarrow \mathbb{R}_+$ and the cost between costs  $d: \mathbb{R}_+ \times \mathbb{R}_+ \rightarrow \mathbb{R}_+$. Whenever appropriate, we will make use of the following, more concise tensor notation. Let $\mathcal{D}$ be a 4-dimensional cost tensor with elements $\mathcal{D}_{ijkl}=d(c_{\mathcal{X}}(x_i, x_k), c_{\mathcal{Y}}(y_j, y_l))$. Define the tensor-matrix multiplication $\mathcal{D}\otimes T \in \mathbb{R}^{n \times m}$ as $[\mathcal{D}\otimes T]_{ij} := \sum_{k,l} \mathcal{D}_{ijkl} T_{kl}$. Then, the GWOT problem can be rewritten as:
\begin{restatable}{eq}{gwotcost}
    \begin{equation}
\mathcal{GW}(\mu, \nu) = \min_{T\in\mathcal{C}_{p,q}}{\langle \mathcal{D}\otimes T, T \rangle}.
\label{eq:GW}\tag{GWOT}
    \end{equation}
\end{restatable}
One important consideration in estimating OT-based distances is the computational burden. \citet{Cuturi2013-xn} showed clear benefits of regularizing the objective function of the OT problem in \eqref{eq:OT} with the negative entropy of the coupling $-\epsilon H(T)$, where
\( H(T) = -\sum_{i,j}{T_{ij}\log(T_{ij}-1)}, \)
and $\epsilon$ is a scaling factor. This not only yields accurate approximations of OT distances when the data include noise but also significantly enhances computational efficiency. Indeed, the entropic-regularized OT (EOT) problem can be solved with a linear convergence rate algorithm, making it efficient for large-scale problems. This result is described in detail in \cref{app:eot}.

A follow-up work from~\citet{Peyre2016-jg} showed that entropic-regularized GWOT (EGWOT) reduces to the problem of solving a sequence of EOT problems, and lowered the time for calculating the GW cost for a fixed coupling from $O(n^2m^2)$ to $O(n^2m + m^2n)$ for the following form of cost functions:
\begin{restatable}{eq}{costclass}
\begin{align}
    d(a, b) = f_1(a) + f_2(b) - h_1(a)h_2(b). \label{eq:cost_class}
\end{align}
\end{restatable}
This includes a wide range of cost functions used in practice, such as the Euclidean distance (Remark 2 of \citet{Peyre2016-jg}.

COOT \citep{Redko2020-bn} is an alternative formulation of the GWOT problem, jointly optimizing the transport of the sample measure $\mu$ onto $\nu$ via coupling $T^s$, and the transport of the feature measures $\beta=\sum_{k=1}^{d_1} r_k\delta_{v_k}, \omega =\sum_{l=1}^{d_2} t_l\delta_{w_l}$, via coupling $T^v$. The Entropic COOT (ECOOT) problem is defined as:
\begin{restatable}{eq}{coot}
\begin{align}
\min_{\substack{T^s\in\mathcal{C}_{p,q}\\ T^v \in \mathcal{C}_{r,t} }}\langle \mathcal{K} \otimes T^v, T^s \rangle - \Omega(T^s, T^v), \label{eq:ecoot}\tag{ECOOT}
\end{align}
\end{restatable}
where $\mathcal{K} \in \mathbb{R}^{n \times m \times d_1 \times d_2}$ is a 4-dimensional cost tensor with elements $\mathcal{K}_{ijkl} = d(x_{ik}, y_{jl})$, $x_{ik}$ and $y_{jl}$ denote the $k$-th and $l$-th feature of the $i$-th sample $x_i$ and the $j$-th sample $y_j$, respectively. The regularization term $\Omega(T^s, T^v)$ is explicitly defined as:
$\Omega(T^s, T^v) = \epsilon^s H(T^s) + \epsilon^v H(T^v)$
where $\epsilon^s$ and $\epsilon^v$ denote the regularizer weights for sample and feature transport respectively.

EOT, EGWOT, and ECOOT have been applied to single-cell perturbation response prediction \citep{Bunne2023-jf, Gu2022-ab} and multiomic integration problems \citep{Demetci2022-qi, Demetci2022-xe, jiang2024scpram,Cao2022}. However, none of these methods, as currently formulated, can leverage additional labels for inference of optimal couplings.

\subsection{Optimal Transport with Additional Structure}
Several works studied the OT problem with additional constraints on the coupling. Structured OT \citep{Alvarez-Melis2018-gw} considered the problem where labeled source samples are transported to unlabeled target samples. \citet{Alvarez-Melis2020-jv} studied the OT problem where source and target samples are independently labeled, and calculated the pairwise distances between the source labels and target labels to improve the coupling between the samples. InfoOT \citep{Chuang2023-ws} promotes the conservation of structure between the source and target space by maximizing the mutual information of the coupling matrix (treated as a joint distribution). HHOT \citep{Yeaton2022-ed} tackles a hierarchical OT problem, where samples' Wasserstein distances are used as the cost to calculate the sample-group-level Wasserstein distances. While these studies solved variants of the OT problem that incorporate some form of label information, they do not address the challenges of matching across different spaces. One significant previous work relevant to ours is \citet{Gu2022-ab}, where a small set of samples are paired (referred to as ``keypoints''). The OT problem is then constrained to include this information. Our setting may be framed as a generalization of their framework, because we only have access to weakly-paired samples, where the pairing is done via a label. 

\subsection{Domain Adaptation via Representation Matching} \label{dann}
Domain adaptation techniques are crucial for overcoming discrepancies between different data distributions. The framework of domain-adversarial neural network (DANN) \citep{Ganin2016-yn} included an adversarial domain classifier to achieve better generalization to unseen data. DANNs have been applied on latent spaces of auto-encoders for single-cell modality integration tasks \citep{Lopez2019-if}, as well as histology and RNA-seq data integration~\citep{Comiter2023-qa}. \citet{Gossi2023-au} estimates OT couplings between latent embeddings of two modalities of single-cell data, solving separate OT problems for each label. Although close to our problem, those methods require a ground truth matching between samples for training which is not available in our scenario. 
JDOT \citep{Courty2017-or} learns the domain-adaptated classifier $f:\mathcal{X} \rightarrow \mathcal{Y}$ by minimizing OT between sample and label pairs of source and target domains $(X_s, Y_s)$ and $(X_t, f(X_t))$. DeepJDOT \citep{Damodaran2018-ka} extends JDOT to learn a classifier for the target domain given a label only available in the source domain, where the training involves OT coupling of latent target and source samples. While these methods are designed for cross-modality prediction, they do not utilize group-level source and target sample matching.

\section{METHOD}

As noted above, GWOT and COOT are suitable computational frameworks for aligning data across modalities but are not designed to leverage labels from perturbation data. The application of GWOT / COOT as-is to our data would mean either learning a global model but ignoring labels altogether, or learning a separate model per label and losing information about the global topology of the phenotypic space. We thus aimed to generalize both the GWOT and COOT problem formulations to incorporate label information. 

\subsection{Notation}

We introduce the following notation to describe our method and data structure concisely. In perturbation data, we observe labels $l^x = \{l^x_{1}, \ldots,l^x_{n}\}$ and $l^y = \{l^y_{1}, \ldots,l^y_{m}\}$ from modalities $\mathcal{X}$ and $\mathcal{Y}$, respectively. Each individual label $l^x_{i}$ and $l^y_{j}$ encodes a perturbation in $\{1, \ldots, L\}$, with $L$ the total number of perturbations. We additionally use the notation $l_x\inv (a) = \{i \mid l^x_i = a\}$ to indicate all the indices of samples in modality $\mathcal{X}$ that undergo perturbation $a \in \{1, \ldots, L\}$. The number of such samples is noted as $n^a = |l_x\inv (a)|$. We define analogously $l_y\inv (a)$, the indices of samples that undergo perturbation $a$, and $m^{a}$, the number of such samples, for modality $\mathcal{Y}$. 

For a general matrix $A$, let $[A]_{\{i_1, \ldots, i_n\}, \{j_1, \ldots, j_m\}}$ denote the submatrix of $A$ with $i_1, \ldots, i_n$-th rows and $j_1, \ldots, j_m$-th columns. For vector $v$, let $[v]_{\{i_1, \ldots, i_n\}}$ denote the subvector of $v$ with $i_1, \ldots, i_n$-th elements. This notation will be used to select specific subsets of our data in subsequent equations.

\subsection{Labeled Entropic-regularized GWOT}
\subsubsection{Definition}
Let $B^l$ be the label-identity matrix defined as $B^l_{ij} :=\mathds{1}\{l^x_{i}=l^y_{j}\}$. We say a coupling $T \in \mathcal{C}_{p,q}$ is $l$-compatible if for all indices $i, j$, we have that:
\begin{align}
    T_{ij} > 0 \implies B^l_{ij}=1,
\end{align} and denote as $\mathcal{C}_{p,q}^l$ the subset of couplings in $\mathcal{C}_{p,q}$ which are $l$-compatible. The \textit{Labeled} EGWOT problem is defined as the EGWOT problem with the additional constraint that $T \in \mathcal{C}_{p,q}^l$. 
We first characterize the structure of the solution of the (simpler) EOT problem with the additional label constraint:
\begin{restatable}{lemma}{leot}\label{lem:leot} 
For a label-identity matrix $B^l$, the $l$-compatible entropic optimal transport plan 
\begin{align}
\mathcal{T}_\epsilon^l(c, p, q) = \argmin_{T \in \mathcal{C}^l_{p, q}}\langle C, T \rangle - \epsilon H(T),
\end{align}
can be expressed as $\textrm{diag}(u)(e^{-C/\epsilon}\odot B^l) \textrm{diag}(v)$, where $\odot$ denotes element-wise multiplication.
\end{restatable}
The proof appears in \cref{app:proof}. This result is important, as it entails that the Labeled EOT problem can be efficiently solved with the celebrated Sinkhorn iterations. 

\subsubsection{Algorithm}

We note that the reduction of the EGWOT problem to an iterative EOT problem, as described in \citet{Peyre2016-jg}, still holds with the additional constraint of $l$-compatible couplings. Therefore, we have the following corollary:
\begin{corollary}
\label{cor:egwl}
For a label-identity matrix $B^l$, the Labeled EGWOT problem can be solved by the iterative update of $T$ with $T^{k+1} \leftarrow \mathcal{T}_\epsilon^l(\mathcal{D} \otimes T^k, p, q)$. 
\end{corollary}
This corollary suggests \cref{alg:egwl} for solving labeled EGWOT problems. In \cref{alg:egwl}, we use the notation $M_{ij} = c_{\mathcal{X}}(x_i, x_j)$ and $\bar{M}_{ij} = c_{\mathcal{Y}}(y_i, y_j)$ to indicate the cost matrices of each modalities. Our implementation relies on the Python library `\verb+ott-jax+' \citep{Cuturi2022-dz} and is provided in the Supplementary Materials.

\begin{algorithm}[t]
   \caption{Computation of $l$-compatible EGWOT}
   \label{alg:egwl}
\begin{algorithmic}[1]
   \STATE {\bfseries Input:} $l^x, l^y, M, \bar{M}, \epsilon, B^l, p, q$
   \STATE Initialize $T$.
   \REPEAT
        \STATE// compute $\mathcal{D} \otimes T$ as in \eqref{eq:cost_per_label} of~\cref{app:cost}.
       \STATE $C_s \leftarrow \mathcal{D} \otimes T$. // is function of $M, \bar{M}$. \label{lst:line:cost}
       \STATE// Sinkhorn iterations to compute $\mathcal{T}^l_\epsilon(C_s, p, q)$
       \STATE Initialize $a \leftarrow \mathds{1}$, set $K \leftarrow e^{-C_s/\epsilon}\odot B^l$.
       \REPEAT
       \STATE $b \leftarrow \frac{q}{K^\top a}$, $a \leftarrow \frac{p}{Kb}$ \label{lst:line:lot}
       \UNTIL convergence
       \STATE Update $T \leftarrow \textrm{diag}(a)(e^{-C_s/\epsilon}\odot B^l)\textrm{diag}(b)$
   \UNTIL convergence
\end{algorithmic}
\end{algorithm}

\subsubsection{Computational Complexity}
Considering the iterative nature of the method, its computational complexity depends on the number of iterations until convergence and the complexity of each iteration. Because the former is hard to quantify, and may depend on the strength of the entropic regularizer~\citep{Peyre2018-pb, Dvurechensky2018-qd, Li2022-mq, Rioux2023-te}, we provide a complexity analysis of each iteration, comprised of the calculation of the cost function, and then of the Sinkhorn iterations. We assume that the cost function $d$ follows the form of \eqref{eq:cost_class}, and that the number of samples per label is balanced: $n^a = n/L$ and $m^a = m/L$ for all $a$. 

We first examine the time complexity of calculating the cross-modal cost matrix $\mathcal{D}\otimes T$. For label $a$ we denote $M^{aa'} := [M]_{l_x\inv(a), l_x\inv(a')}$ as submatrix of the cost matrix $M$ for modality $\mathcal{X}$ between perturbations $a$ and $a'$. We define $\bar{M}^{aa'}$ analogously for modality $\mathcal{Y}$. Then, for each label $a$, the corresponding block of $\mathcal{D}\otimes T$ is computed as:
\begin{align*} 
&[\mathcal{D}\otimes T]_{ l_x\inv (a),  l_y\inv (a)} = \sum_{a'=1}^L f_1(M^{aa'})p^{a'} \numberthis \label{eq:cost_per_label} \\&+  \sum_{a'=1}^L f_1(\bar{M}^{aa'})q^{a'} - \sum_{a'=1}^L h_1(M^{aa'})T^{aa'}h_2(\bar{M}^{aa'})^\top,
\end{align*}
where $p^{a'}=[p]_{l_x\inv(a')}$ and $q^{a'}=[q]_{l_y\inv(a')}$. The proof appears in~\cref{app:cost}. From this result follows the time complexity:
\begin{remark} \label{rem:lgw_complexity}
 Calculation of the cross-modal cost matrix $\mathcal{D}\otimes T$ has time complexity $O((n^2m + nm^2)/L)$. This represents an L-fold improvement over the $O(n^2m + nm^2)$ complexity of standard (non-labeled) GWOT.
\end{remark}
We can also accelerate the calculation of the Sinkhorn interations by a factor of $L$ using block-level updates (further details in~\cref{app:alg}).
\begin{remark} \label{rem:lot_complexity}
The time complexity of one block-level Sinkhorn interation is $O(nm/L)$. This represents an L-fold improvement over the $O(nm)$ complexity of a standard Sinkhorn iteration.
\end{remark}
The labeled version of GWOT therefore has an overall $L$ times speedup compared to the vanilla implementation. These accelerations are advantageous, as a large screen dataset may be comprised of thousands of perturbations, with a fixed (small) number of cells per perturbation.




\subsection{Labeled COOT}

We propose an efficient algorithm for \textit{Labeled} versions of both the COOT and ECOOT problem, based on the iterative procedure proposed by \citet{Redko2020-bn}. Briefly, we posit for each label $a$ sample transport matrix ${T^s}^{(a)}$, as well as a shared global feature transport $T^v$. At each iteration, we update each of the $L$ sample transports ${T^s}^{(a)}$ for each label $a$ and the shared global feature transport $T^v$. 

\subsubsection{Definition}
For the $l$-compatible matrix $T^s$, we define ${T^s}^{(a)} := [T^s]_{l_x\inv(a), l_y\inv(a)}$, the restriction of the transport to samples with label $a$. Based on the observation that ${T^s}^{(a)}\in \mathcal{C}_{p^a q^a}$, where $p^a=[p]_{l_x\inv(a)}$ and $q^a=[q]_{l_y\inv(a)}$, we may define the following Labeled ECOOT problem:
\begin{equation}
\min_{\substack{T^v \in \mathcal{C}_{r,t} \\ \{{T^s}^{(a)}\in \mathcal{C}_{p^aq^a}\}^L_{a=1}} } \sum_{a = 1}^L  \langle \mathcal{K}^{(a)} \otimes T^v, {T^s}^{(a)} \rangle - \Omega({T^s}^{(a)}, T^v), 
\label{eq:ecootl}\tag{Labeled ECOOT}
\end{equation}
where $\mathcal{K}^{(a)} \in \mathbb{R}^{n^a \times m^a \times d_1 \times d_2}$ is the 4-dimensional cost tensor for label $a$, with elements $\mathcal{K}^{(a)}_{ijkl} = d(x_{ik}, y_{jl})$ for $(i, j) \in l_x\inv(a) \times l_y\inv(a)$. The regularization term $\Omega({T^s}^{(a)}, T^v) = {\epsilon^s}^{(a)}H({T^s}^{(a)}) + \epsilon^v H(T^v)$, where ${\epsilon^s}^{(a)}$ and $\epsilon^v$ are the regularizer weights for sample and feature transport respectively.

\begin{algorithm}[t]
   \caption{BCD algorithm for Labeled ECOOT}
   \label{alg:ecootl}
\begin{algorithmic}[1]
   \STATE {\bfseries Input:} $\mathcal{K}, \mathcal{K}', l, r, t, p, q, \epsilon^v, \epsilon^{s(1)},\ldots, \epsilon^{s(L)}$
   \STATE Initialize $T^{s(1)}$, \ldots, $T^{s(L)}$, $T^v$.
   \REPEAT
        \STATE $T^{v} \leftarrow \mathcal{T}_{\epsilon^v}(r, t, \sum_{a=1}^L\mathcal{K}'^a\otimes {T^s}^{(a)})$.
        \FOR{$a=1$ {\bfseries to} $L$}
            \STATE ${T^s}^{(a)} \leftarrow \mathcal{T}_{{\epsilon^s}^{(a)}}(p^a, q^a, \mathcal{K}^a \otimes T^v)$.
       \ENDFOR
   \UNTIL convergence
\end{algorithmic}
\end{algorithm}

\subsubsection{Algorithm} 
Recall $\mathcal{K}$ denotes the 4-way tensor with elements $\mathcal{K}_{ijkl} = d(x_{ik}, y_{jl})$. We denote $\mathcal{K}'$ as the image of $\mathcal{D}$ by a transposition, with elements $\mathcal{K}'_{klij} = d(x_{ik}, y_{jl})$. For each label $a$, $\mathcal{K}'^{(a)}$ denotes a subtensor of $\mathcal{K}'$ with elements $\mathcal{K}'^{(a)}_{klij}$ for $(i, j) \in l_x\inv(a) \times l_y\inv(a)$. Finally, $\mathcal{T}_\epsilon(p, q, c)$ denotes the entropic OT solution obtained by the Sinkhorn algorithm.

The block coordinate descent (BCD) algorithm in algorithm 1 of \citet{Redko2020-bn} can still be accurately employed for this problem. Indeed, independent updates of ${T^s}^{(1)}, \ldots, {T^s}^{(L)}$ together are equivalent to a single step of the sample transport update. We adapted the BCD algorithm for labeled COOT as \cref{alg:ecootl}, whose line 4 is derived in \cref{app:cootl}. Our implementation, based on `\verb+ott-jax+' \citep{Cuturi2022-dz}, is provided in the Supplementary Materials.  

\subsubsection{Computational Complexity}
The computational complexity of COOT for each iteration consists of the complexity for the sample-to-sample OT problem and the feature-to-feature OT problem. The Labeled COOT framework provides a $L$ times speedup for solving the sample-to-sample OT problem, but has unchanged complexity for the feature-to-feature OT problem. In our experiment, where the number of samples $n, m$ is significantly larger than the number of dimensions of each modality $d_1, d_2$, the former problem dominates asymptotically, resulting in an overall speedup of a factor $L$. We provide further details, as well as derivations of the space complexity in \cref{app:cootl}.

\section{EXPERIMENTS}

We tested whether we could (1) match samples across modalities so that similar samples are matched with each other and (2) predict the response to perturbations in one modality given the response to that perturbation measured in the other modality. We further (3) obtained feature matching by plugging the learned sample matching $T_s$ into \ref{eq:ecoot} and solving the feature-to-feature OT problem with the fixed $T_s$. Full experimental details appear in \cref{app:exp}.

\subsection{Data}
\begin{figure}[h!]
\centering
\vspace{-0.1in}
\includegraphics[width=0.95\columnwidth]{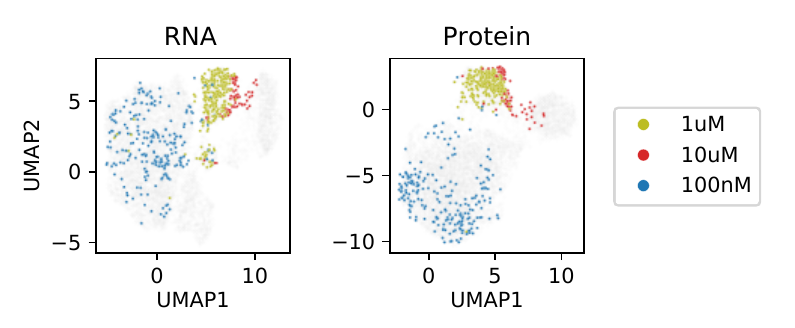}
\vspace{-0.1in}
\caption{UMAPs of cells treated with one of the kinase inhibitors (AZD1480) with varying dosages}
\vspace{-0.1in}
\label{fig:azd_umap}
\end{figure}

The dataset records single-cell RNA (2000 genes) and protein (123 proteins) profiles of T cells undergoing T cell receptor (TCR) activation following perturbation with kinase inhibitors for multimodal understanding of TCR activation. The dataset includes 11 inhibitors, used at varying dosages (100 nM, 1 {\textmu}M, and 10 {\textmu}M), as well as negative controls (vehicle and non-activation) (Figure \ref{fig:azd_umap}). We normalized and scaled features following conventional single-cell data processing pipelines \citep{Wolf2018-gl} and used the first 50 principal components of each modality as the input for OT-based methods. 

\begin{figure*}[t]
\centering
\vspace{-0.2in}
\includegraphics[width=1.8\columnwidth,height=4cm]{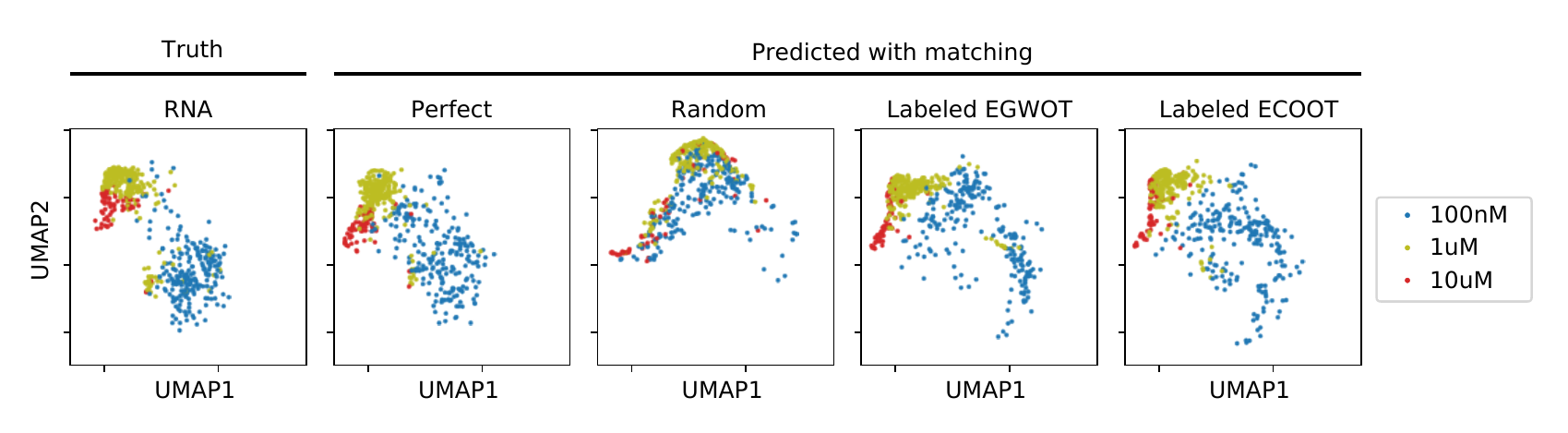}
\vspace{-0.2in}
\caption{UMAPs of predicted RNA modality of AZD1480 with varying dosages.}
\vspace{-0.1in}
\label{fig:azd_pred}
\end{figure*}

We also provide supplementary results on synthetic data, generated from a hierarchical Bayesian model in appendix~\ref{app:simulation}

\subsection{Methods}
We compared labeled ECOOT and EGWOT against the baseline of EOT, ECOOT, and EGWOT without sample labels (\textit{no label}), and per label EOT and EGWOT. For the prediction task, we trained a multilayer perceptron (MLP) to predict RNA levels from protein measurements according to the learned coupling $T$. Specifically, for training, a sample from the RNA modality $y_j$ is randomly sampled for each sample from the protein modality $x_i$ as $j \sim \text{Multinomial}(T_{i\cdot}/\sum T_{i\cdot})$, independently for each minibatch. For the matching and prediction task, we included additional baselines of domain-adversarial VAE (DAVAE) and its naive label adaptation described in \cref{app:davae}. All experiments were conducted using an internal computing cluster with 10 CPU cores and 10GB of RAM per CPU.

\subsection{Evaluation}
We evaluated the matching for the observed treatments and assessed the prediction for the held-out treatments. Because the data were collected using a multi-modal profiling method (where RNA and protein profiles are measured jointly for each single cell), we have a ground-truth cell-to-cell matching from the data. Thus, for the matching, we calculated the barycentric FOSCTTM \citep{Liu2019-zj}, that is the fraction of the barycentric projections closer than the true match as in \citet{Demetci2022-qi}. We further evaluated the matching using the ground-truth treatment dosages in RNA profiles ($\{d^x_{1}, \ldots, d^x_{n}\}$) and protein profiles ($\{d^y_{1}, \ldots, d^y_{m}\}$). Given the treated cell profiles, we expect cells treated with the same dose to match. We thus use the dose labels to evaluate the matching, by taking the sum of the coupling weights $\sum_{ij} T_{ij}$ for the samples with matching doses $(i, j) \in \{i,j \mid l^x_{i}=l^y_{j}=k, d^x_{i}=d^y_{j}\}$ and refer to it as the \textit{dosage match} metric. 
For the prediction task, we calculated the mean Pearson ($R$) and Spearman ($\rho$) correlation coefficients between gene expression fold changes in real vs. predicted profiles, where the fold changes are calculated over the mean expression level of genes in the cells treated by vehicle. This was done for each cell ($R_v$, $\rho_v$), and between cells for each feature ($R_s$, $\rho_s$). We also report the mean squared error (MSE) between the predicted and true gene expression profile. For the feature matching, we calculated the enrichment of coupling weights in 23 ground truth protein and RNA matches over the uniform feature coupling (\cref{app:exp}). Metrics are calculated for the ground truth one-to-one matching (``Perfect''), uniform matching of cells with the same dosage (``By dosage'', when $T_{ij} = c$ for $(i, j) \in \{i,j \mid l^x_{i}=l^y_{j}, d^x_{i}=d^y_{j}\}$ and otherwise $0$ with some constant $c$), and uniform matching with the same perturbation labels (``Uniform per label'', $T_{ij} = c$ for $(i, j) \in \{i,j \mid l^x_{i}=l^y_{j}\}$ and otherwise $0$ with some constant $c$).

\begin{table}[h]
\vspace{-0.15in}
\caption{Evaluation metrics of OT and GW approaches for sample matching task.}
\vspace{-0.1in}
\label{tbl:metrics_matching}
\begin{center}
\resizebox{\columnwidth}{!}{%
\begin{tabular}{rlccc}
\toprule
\multicolumn{2}{c}{Method} & \makecell{Bary\\FOSTTM($\downarrow$)} & \makecell{Dosage\\match ($\uparrow$)}& \makecell{Mean\\rank} \\
\midrule
\multicolumn{2}{c}{Perfect}& 0 & 1 & -\\
\multicolumn{2}{c}{By dosage}& 0.239 & 1 & -\\
\multicolumn{2}{c}{Uniform per label}& 0.298 & 0.357 & -\\
\midrule
EOT     & no label & 0.428 & 0.040 & 10 \\
        & per label     & 0.336 & 0.346 & 5.5 \\
ECOOT   & no label & 0.414 & 0.049 & 9 \\
        & per label     & 0.359 & 0.380 & 5.5 \\
        & labeled       & 0.270 & \textbf{0.456} &  2 \\
EGWOT   & no label & 0.373 & 0.068 & 8 \\
        & per label     & 0.332 & 0.381 & 4 \\
        & labeled       & 0.283 & \underline{0.452} & 3 \\
DAVAE   & no label & \textbf{0.231} & 0.206 & 3.5    \\
        & labeled       & \underline{0.242} & 0.205 & 4.5  \\
\bottomrule
\end{tabular}}
\end{center}
\vspace{-0.15in}
\end{table}

\begin{table*}[h]
\caption{Evaluation metrics of OT and GW approaches for prediction task.}
\label{tbl:metrics_prediction}
\begin{center}
\begin{tabular}{rlcccccc}
\toprule
\multicolumn{2}{c}{Method} & $R_v$ ($\uparrow$) & $\rho_v$ ($\uparrow$) & $R_s$ ($\uparrow$) & $\rho_s$ ($\uparrow$) & MSE ($\downarrow$) & \makecell{Mean\\rank} \\
\midrule
\multicolumn{2}{c}{Perfect}& 0.107 & 0.118 & 0.163 & 0.149 & 0.258& - \\
\multicolumn{2}{c}{By dosage}& 0.0812 & 0.0448 & 0.0903 & 0.0863 & 0.264& - \\
\multicolumn{2}{c}{Uniform per label}& 0.0794 & 0.0403 & 0.0761 & 0.0781 & 0.264& - \\
\midrule
EOT     & no label & 0.0482 & 0.007 & 0.0068 & 0.0063 & 0.287& 8 \\
        & per label     & 0.0544 & 0.0239 & 0.0345 & 0.0307 & 0.283& 6.2 \\
ECOOT   & no label & 0.053 & 0.0207 & 0.0395 & 0.0408 & 0.282& 5.8 \\
        & per label     & 0.0806 & 0.0443 & 0.0776 & \underline{0.0815} & 0.265 & 2.6 \\
        & labeled       & \textbf{0.0852}  & \textbf{0.0523} & \textbf{0.0854} & 0.0778 & \underline{0.265} & 1.6 \\
EGWOT   & no label & 0.0631 & 0.0227 & 0.0302 & 0.034 & 0.282 & 5.8 \\
        & per label     & 0.0785 & \underline{0.0449} & 0.0737 & 0.0737 & 0.265 & 3.2 \\
        & labeled       & \underline{0.0836} & 0.044 & \textbf{0.0854} & \textbf{0.0825} & \textbf{0.264} & 1.8 \\
DAVAE   & no label & 0.0342 & -0.0069 & 0.0006 & -0.0001 & 0.33 & 9 \\
        & labeled       & 0.0182 & -0.0079 & -0.0016 & -0.0014 & 0.332 & 10 \\
\bottomrule
\end{tabular}
\end{center}
\end{table*}

\begin{table}[h]
\vspace{-0.1in}
\caption{Evaluation metrics of OT and GW approaches for feature matching task.}
\label{tbl:metrics_feature}
\begin{center}
\begin{tabular}{rlcc}
\toprule
\multicolumn{2}{c}{Method} & \makecell{Enrich-\\ment($\uparrow$)} & Rank \\
\midrule
\multicolumn{2}{c}{Perfect}& 6.95 & - \\
\multicolumn{2}{c}{By dosage}& 5.16 & - \\
\multicolumn{2}{c}{Uniform per label}& 1.85 & - \\
\midrule
EOT     & no label & 1.10 & 7 \\
        & per label     & 1.26 & 5 \\
ECOOT   & no label & 1.07 & 9 \\
        & per label     & 1.97 & 4 \\
        & labeled       & \underline{5.31} & 2 \\
EGWOT   & no label & 3.74 & 3 \\
        & per label     & 1.26 & 5 \\
        & labeled       & \textbf{19.8} & 1 \\
DAVAE   & no label & 1.04 & 10 \\
        & labeled       & 1.09 & 8 \\
\bottomrule
\end{tabular}
\end{center}
\vspace{-0.1in}
\end{table}

\subsection{Hyperparameter selection}
We conducted a nested 5-fold cross-validation (CV) by splitting treatments into a train, validation, and test sets. The best hyperparameters for matching and prediction tasks were independently selected from the inner CV. We performed a hyperparameter search for the entropic regularizer weight $\epsilon \in \{10^{-2}, 10^{-3}, 10^{-4}, 10^{-5}\}$ where the maximum cost was normalized as $1$ for entropic OT and GW methods. COOT methods used the same $\epsilon$ for all sample and feature OTs. The scale of the adversarial loss $\lambda^{Adv}$ was optimized across $\{1, 5, 10, 50, 100\}$ for DAVAEs. Hyperparameters with the least barycentric FOSCTTM for the matching and the highest $R_s$ for the prediction were independently selected for each outer CV fold. We report the mean evaluation metrics across the outer folds.
For feature matching, we obtained the best sample matchings with the highest dosage match among the $\epsilon$ in the same grid to calculate feature matchings across the same $\epsilon$'s and report the best enrichment. 

\subsection{Results}

Overall, GWOT-based methods (EGWOT and ECOOT variants) outperformed OT-based methods (\cref{tbl:metrics_matching,tbl:metrics_prediction,tbl:metrics_feature}). This highlights the importance of optimizing \textit{distance between distances} rather than naively assuming the existence of shared cost metric even when the number of the dimensions of the source and target modalities are the same.

DAVAE methods showed poor prediction performance despite decent matching (\cref{tbl:metrics_matching,tbl:metrics_prediction}). This may be due to the instability of adversarial training \citep{Kodali2017-qi}, the small hyperparameter search space, and the inherent tradeoff between the matching and prediction of DANNs. A comprehensive hyperparameter search and alternative domain adaptation methods may improve the performance.

Within GWOT-based methods, not using the label information led to poor performance, as they fail to harness the strong matching information provided by the input sample labels. Per-label GWOT-based methods performed better than methods without label input, but did not achieve as high matching and prediction performance as the label-aware GWOT-based methods. 

The superior sample matching, prediction, and feature matching performance of labeled GWOT-based methods stems from information sharing across different labels, while still constraining the sample matching by labels. Specifically, when Labeled EGWOT calculates the cost between samples with the same label, it will consider the distance of each sample to the samples with \textit{other} labels (\cref{app:cost}). For Labeled ECOOT, the sample couplings for all labels explicitly share the global coupling between features. We present further results including standard deviations of evaluation metrics and visualizations in \cref{app:result}.

\section{DISCUSSION}
We provide a mathematical and algorithmic adaptation of GWOT methods for the case when sample labels are available. These methods outperform OT, GWOT, and DAVAE baselines both in cross-modality matching and prediction tasks.
Labeled GWOT-based methods had improved matching between raw features from the sample matching (\cref{tbl:metrics_feature}), suggesting their potential to improve the interpretability of features. We note that learning sample matching from the latent representation provides substantial computational acceleration over the learning in the raw feature space. 


We note that application of our approach requires the removal of modality-specific noise that may dominate the cost calculation. As GWOT methods fundamentally rely on the similarity of cost distributions across modalities, depending on the datasets, their application may require the preprocessing procedure or latent representations that would seek for the common variance and remove technical artifacts.

In this work, we provide the extensions of \citet{Peyre2016-jg} and \citet{Redko2020-bn} for the datasets with labeled samples and show they provide the acceleration by the number of labels. For future work, we wish to explore recent methodology aiming at scaling up calculations of GWOT plans. Low-rank decomposed cost functions \citep{Scetbon2022-ng} provides substantial acceleration. Neural GW approaches would likely provide better scalability \citep{Carrasco2023-jw, Wang2023-zc} and flexibility beyond specific forms of cost function \citep{Klein2023-me}. These methods harbor much potential to be adapted to the labeled datasets and provide further acceleration and generalizability.

We expect our framework to combine the modality-specific strengths of different high content perturbation screens, such as the high scalability of optical pooled screens \citep{Feldman2019-vb} and the high resolution and interpretability of Perturb-seq \citep{Dixit2016-kw}. As GWOT-based methods rely on the sample-to-sample cost within each modality, more sophisticated latent representation may be needed to remove any large modality-specific variations.

\subsection*{Code Availability Statement}
We implement our new model and benchmarks using the Python `\verb+scvi-tools+' ~\citep{gayoso2022python} and `\verb+ott-jax+' ~\citep{Cuturi2022-dz} libraries, and release it as open-source software at \url{https://genentech.
github.io/Perturb-OT/}. 

\subsection*{Data Availability Statement} Input data used for the experiments in this manuscript are currently undergoing publication as a separate manuscript. Data and software are attached as the supplementary material.

\subsection*{Acknowledgments and Funding Disclosures}
We thank Kelvin Chen for data generation, as well as discussions about the dataset. We thank Takamasa Kudo for early
discussions that helped frame the problem. We acknowledge
Jan-Christian Huetter for discussions about the optimal transport literature, as well as Alma Andersson and Aicha Bentaieb for discussions about software implementation.

Disclosures: This work was done while Jayoung Ryu was
an intern at Genentech. Romain Lopez, Charlotte Bunne,
and Aviv Regev are employees of Genentech. Aviv Regev is a co-founder and equity holder of Celsius Therapeutics
and an equity holder in Immunitas and Roche.


\bibliography{references}

\clearpage
\section*{Checklist}

 \begin{enumerate}

 \item For all models and algorithms presented, check if you include:
 \begin{enumerate}
   \item A clear description of the mathematical setting, assumptions, algorithm, and/or model. [Yes]
   \item An analysis of the properties and complexity (time, space, sample size) of any algorithm. [Yes]
   \item (Optional) Anonymized source code, with specification of all dependencies, including external libraries. [Yes]
 \end{enumerate}

 \item For any theoretical claim, check if you include:
 \begin{enumerate}
   \item Statements of the full set of assumptions of all theoretical results. [Yes]
   \item Complete proofs of all theoretical results. [Yes]
   \item Clear explanations of any assumptions. [Yes]     
 \end{enumerate}

 \item For all figures and tables that present empirical results, check if you include:
 \begin{enumerate}
   \item The code, data, and instructions needed to reproduce the main experimental results (either in the supplemental material or as a URL). [Yes, except the data that will be published separately.] 
   \item All the training details (e.g., data splits, hyperparameters, how they were chosen). [Yes]
         \item A clear definition of the specific measure or statistics and error bars (e.g., with respect to the random seed after running experiments multiple times). [Yes]
         \item A description of the computing infrastructure used. (e.g., type of GPUs, internal cluster, or cloud provider). [Yes]
 \end{enumerate}

 \item If you are using existing assets (e.g., code, data, models) or curating/releasing new assets, check if you include:
 \begin{enumerate}
   \item Citations of the creator If your work uses existing assets. [Yes]
   \item The license information of the assets, if applicable. [Yes]
   \item New assets either in the supplemental material or as a URL, if applicable. [Yes]
   \item Information about consent from data providers/curators. [Not Applicable]
   \item Discussion of sensible content if applicable, e.g., personally identifiable information or offensive content. [Not Applicable]
 \end{enumerate}

 \item If you used crowdsourcing or conducted research with human subjects, check if you include:
 \begin{enumerate}
   \item The full text of instructions given to participants and screenshots. [Not Applicable]
   \item Descriptions of potential participant risks, with links to Institutional Review Board (IRB) approvals if applicable. [Not Applicable]
   \item The estimated hourly wage paid to participants and the total amount spent on participant compensation. [Not Applicable]
 \end{enumerate}

 \end{enumerate}

\newpage
\appendix
\onecolumn
\section*{Appendices}
This appendix provides detailed information on various aspects of our work. The following sections contain:
\begin{enumerate}
    \item Introduction of entropy-regularized OT is described in \cref{app:eot}.
    \item The proof of Lemma \ref{lem:leot} is detailed in \cref{app:proof}.
    \item The acceleration of the cost calculation for both unlabeled and labeled GWOT is described in \cref{app:cost}, following the assumptions of~\citet{Peyre2016-jg} regarding the decomposition for the cost function.
    \item The accelerated Sinkhorn updates for solving labeled EGWOT problems are presented in \cref{app:alg}.
    \item The labeled COOT problem and our proposed algorithm for numerical resolution are explained in \cref{app:cootl}.
    \item Details about data pre-processing, model architecture, and training procedures used for the experiments are provided in \cref{app:exp}.
    \item Supplementary results on synthetic data appear in \cref{app:simulation}.
    \item A naive adaptation of DAVAE to the labeled setting is described in \cref{app:davae}.
    \item Extended experimental results are shown in \cref{app:result}.
\end{enumerate}

\section{Entropy-regularized OT} \label{app:eot}
This section introduces the background of entropy-regularized OT. We first show the entropy regularization gives optimization objective that can be solved by Sinkhorn iterations. Then we visualize the coupling matrix through Sinkhorn iteration for a toy example for illustrative purposes. 
\subsection{Entropy-regularized OT can be solved through Sinkhorn iterations}
We restate the result of \citet{Cuturi2013-xn} to introduce the acceleration allowed by adding entropic regularizer. Entropy-reguliarzed optimal transport problem is defined as follows:
\begin{equation}
    \mathcal{OT}_\epsilon(p, q) = \min_{T \in \mathcal{C}_{p,q}} \langle C, T \rangle - \epsilon H(T) \label{eq:EOT}\tag{EOT}
\end{equation}
where \( H(T) = -\sum_{i,j}{T_{ij}\log(T_{ij}-1)}, \). 
We may equivalently reformulate the optimization problem using Lagrange multipliers:
\begin{equation}
    \mathcal{OT}_\epsilon(p, q) = \min\limits_{T}\max\limits_{\lambda^p, \lambda^q} \langle C, T \rangle - \epsilon H(T) + \langle \lambda^p, p - T \mathds{1}_m \rangle + \langle \lambda^q, q - T^\top\mathds{1}_n \rangle \nonumber 
\end{equation}
and then use the dual formulation:
\begin{equation}
    \mathcal{OT}_\epsilon(p, q) = \max\limits_{\lambda^p, \lambda^q} \min\limits_{T} \langle C, T \rangle - \epsilon H(T) + \langle \lambda^p, p - T \mathds{1}_m \rangle + \langle \lambda^q, q - T^\top\mathds{1}_n \rangle  
    \label{eq:ot_dual}
\end{equation}
where strong duality is guaranteed since the objective function is convex in $T$ and the constraints are affine. The solution to the inner minimization problem, for fixed $\{\lambda^p, \lambda^q\}$ is obtained by finding the critical point:
\begin{align}
    \bar{T}_{ij}(\lambda^p, \lambda^q) = \exp\left(\frac{1}{\epsilon}\left(\lambda^p_i + \lambda^q_j - c_{ij}\right)\right).   \label{eq:optimal_t_labels}
\end{align}
Now, we notice that for variable $\bar{T}_{ij}(\lambda^p, \lambda^q)$, we have that the regularized cost is derived as:
\begin{align}
    \langle C, \bar{T} \rangle - \epsilon H(\bar{T}) = \sum_{ij} {C_{ij} \bar{T}_{ij}} + \epsilon \bar{T}_{ij} (\log \bar{T}_{ij} - 1),
\end{align}
which, by injecting the value of $\bar{T}_{ij}(\bm{\Lambda})$ into this expression, we obtain:
\begin{align}
    \langle C, \bar{T} \rangle - \epsilon H(\bar{T}) = \sum_{ij}  {C_{ij} \bar{T}_{ij}} + \bar{T}_{ij} \left(\lambda^p_i + \lambda^q_j - c_{ij} - \epsilon \right),
\end{align}
from which, after developing the products and identifying dot products, we finally obtain the simplified expression:
\begin{align}
    \langle C, \bar{T} \rangle - \epsilon H(\bar{T}) =\langle\lambda^p, \bar{T} \mathds{1}_m \rangle +\langle\lambda^q, \bar{T}^\top \mathds{1}_n \rangle -\epsilon \langle \mathds{1}_n, \bar{T}\mathds{1}_m\rangle.
\end{align}
Finally, we may inject this result into the dual formulation from~\eqref{eq:ot_dual} to obtain 
\begin{align}
    \mathcal{OT}_\epsilon(p, q)
    &= \max\limits_{\lambda^p, \lambda^q} \langle \lambda^p, p \rangle + \langle \lambda^q, q \rangle -\epsilon \langle \mathds{1}_n,  \bar{T}_{ij}(\lambda^p, \lambda^q)\mathds{1}_m\rangle\label{eq:ot_plugged_t}. \\
    &= \max\limits_{\lambda^p, \lambda^q} \langle \lambda^p, p \rangle + \langle \lambda^q, q \rangle -\epsilon \sum_{ij} \bar{T}_{ij} (\lambda^p, \lambda^q) \\
    &= \max\limits_{\lambda^p, \lambda^q} \langle \lambda^p, p \rangle + \langle \lambda^q, q \rangle + \langle R^\epsilon(\lambda^p, \lambda^q, C), \mathds{1}_n \mathds{1}_m^\top \rangle, \label{eq:new_opt0}
\end{align}
where $[R^\epsilon(\lambda^p, \lambda^q, C)]_{ij} = -\epsilon\exp(\frac{1}{\epsilon}(\lambda^p_i + \lambda^q_j - c_{ij}))$. Defining $({\lambda^p}^*, {\lambda^q}^*)$ as the optimal solution for $\eqref{eq:new_opt0}$, we now get the optimal transport plan $T^*$ for \eqref{eq:ot_dual}:
\begin{align}
    T_{ij}^*({\lambda^p}^*, {\lambda^q}^*) &=  \exp(\frac{1}{\epsilon}({\lambda^p_i}^* + {\lambda^q_j}^* - C_{ij})) \\
    &= \textrm{diag}(a)K\textrm{diag}(b) \nonumber
\end{align}
where 
\begin{align}
a_i = \exp\left(\frac{{\lambda^p_i}^*}{\epsilon}\right), \ b_j = \exp\left(\frac{{\lambda^q_j}^*}{\epsilon}\right), \ K_{ij} = \exp\left(-\frac{C_{ij}}{\epsilon}\right).
\end{align}
This can be solved through the following Sinkhorn iterations.
\begin{align}
a_i &\leftarrow \frac{p_i}{\sum_{j=1}^m K_{ij}b_j} \nonumber \\
b_j &\leftarrow \frac{q_j}{\sum_{i=1}^n K_{ji}a_i} \nonumber
\end{align}

\subsection{Visualization of Sinkhorn updates}
\begin{figure}
\centering
\includegraphics[width=0.95\columnwidth]{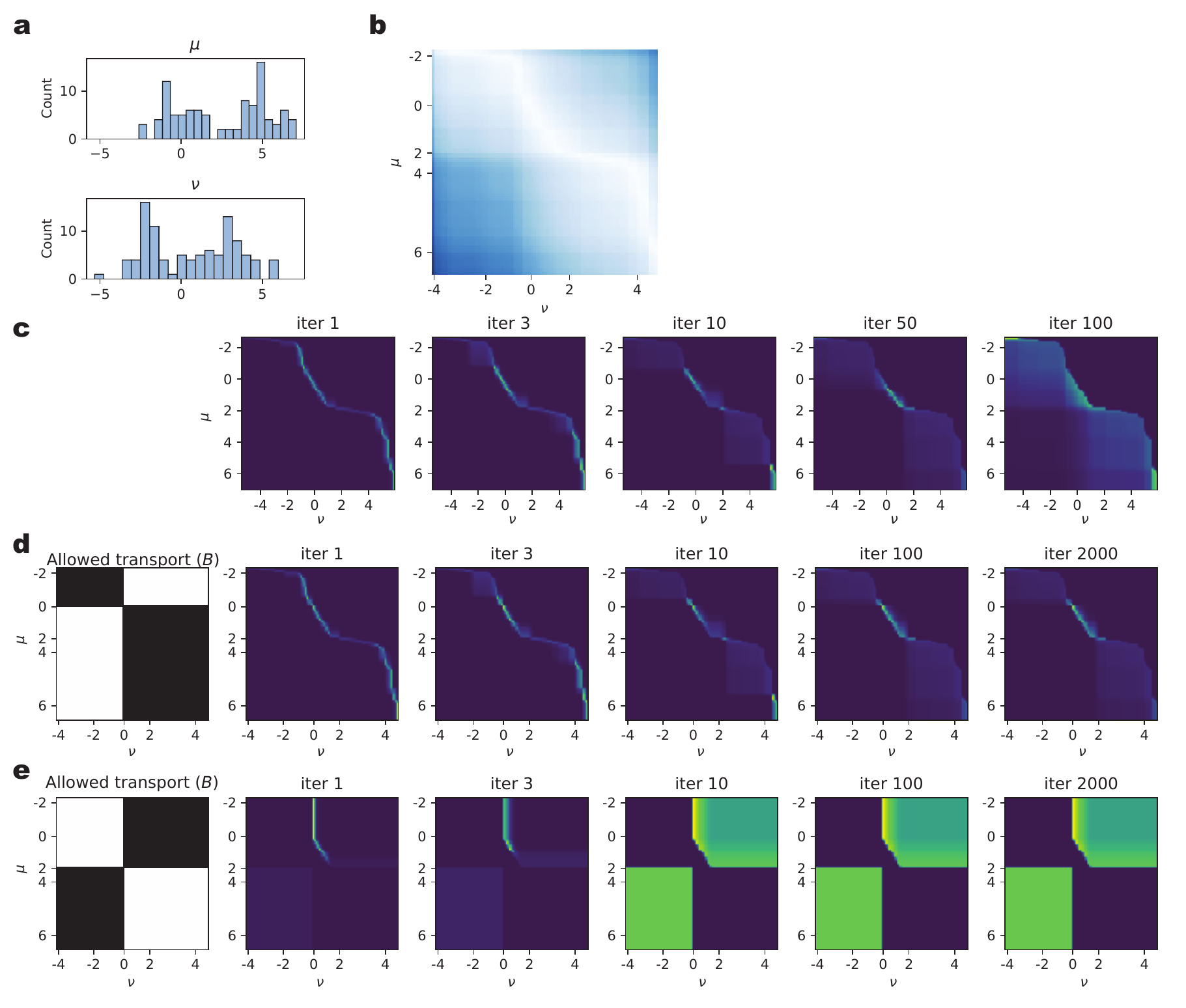}
\caption{Examples of EOT and $l$-compatible EOT coupling matrices through Sinkhorn iterations. (a) Histogram of $x$ and $y$ (b) Cost matrix (c) Coupling matrices along EOT Sinkhorn iterations (b) Coupling matrices along Sinkhorn iterations for $l$-compatible EOT where $l^x=\{\mathbf{1}(x_i>0)\}, l^y=\{\mathbf{1}(y_j>0)\}$ (d) Coupling matrices along Sinkhorn iterations for $l$-compatible EOT where $l^x=\{\mathbf{1}(x_i<2)\}, l^y=\{\mathbf{1}(y_j>0)\}$.} 
\label{illustration}
\end{figure}

We visualize the coupling matrices for EOT and labeled EOT for two example data labels in Figure \ref{illustration}. $\mu$ and $\nu$ are discrete distributions with $n=m=100$, each sampled from Gaussian mixture distributions $\textrm{MixNormal}([0, 5]^\top, I)$ and $\textrm{MixNormal}([-2, 3]^\top, 2I)$. 

\section{Proof of \cref{lem:leot}} \label{app:proof}
We remind the reader that the label-identity matrix $B^l$ is defined as $B^l_{ij} =\mathds{1}\{l^x_{i}=l^y_{j}\}$, where  $l^x = \{l^x_{1}, \ldots,l^x_{n}\}$ and $l^y = \{l^y_{1}, \ldots,l^y_{m}\}$ denotes the perturbation labels from modalities $\mathcal{X}$ and $\mathcal{Y}$, respectively. 
\leot*
We note that comparing \cref{lem:leot} with lemma 2 in \citet{Cuturi2013-xn}, the solution of $l$-compatible OT is equivalent to the regular EOT with $c(x_i, y_j) = +\infty$ for $(i, j) \in \{(i, j) \mid l_{x_i} \neq l_{y_j}\}$. The proof below makes this argument more precise.
\begin{proof}
The $l$-compatible EOT (Labeled EOT) problem is defined as the following optimization problem:  
\begin{equation}
    \mathcal{OT}_\epsilon^l(p, q) = \mathop{\min}\limits_{T \in \mathcal{C}_{p, q}^l}\langle C, T \rangle - \epsilon H(T) \label{eq:LEOT}
\end{equation}
We may equivalently reformulate the optimization problem using Lagrange multipliers:
\begin{align}
    \mathcal{OT}_\epsilon^l(p, q) = \min\limits_{T}\max\limits_{\lambda^p, \lambda^q, \Lambda^B} \langle C, T \rangle - \epsilon H(T) + &\langle \lambda^p, p - T \mathds{1}_m \rangle + \langle \lambda^q, q - T^\top\mathds{1}_n \rangle \nonumber \\
    &+ \langle T\odot (\mathds{1}_n\mathds{1}_m^\top-B^l)), \Lambda^B \rangle
\end{align}
and then use the dual formulation:
\begin{align}
    \mathcal{OT}_\epsilon^l(p, q) = \max\limits_{\lambda^p, \lambda^q, \Lambda^B} \min\limits_{T} \langle C, T \rangle - \epsilon H(T) + &\langle \lambda^p, p - T \mathds{1}_m \rangle + \langle \lambda^q, q - T^\top\mathds{1}_n \rangle \nonumber \\
    &+ \langle T\odot (\mathds{1}_n\mathds{1}_m^\top-B^l), \Lambda^B \rangle, \label{eq:dual}
\end{align}
where strong duality is guaranteed since the objective function is convex in $T$ and the constraints are affine. The solution to the inner minimization problem, for fixed $\bm{\Lambda} = \{\lambda^p, \lambda^q, \Lambda^B\}$ is obtained by finding the critical point:
\begin{align}
    \bar{T}_{ij}(\bm{\Lambda}) = \exp\left(\frac{1}{\epsilon}\left(-(1-B^l_{ij})\Lambda^B_{ij} + \lambda^p_i + \lambda^q_j - c_{ij}\right)\right).   \label{eq:leot_optimal_t_labels}
\end{align}
Now, we notice that for variable $\bar{T}_{ij}(\bm{\Lambda})$, we have that the regularized cost is derived as:
\begin{align}
    \langle C, \bar{T} \rangle - \epsilon H(\bar{T}) = \sum_{ij} {C_{ij} \bar{T}_{ij}} + \epsilon \bar{T}_{ij} (\log \bar{T}_{ij} - 1),
\end{align}
which, by injecting the value of $\bar{T}_{ij}(\bm{\Lambda})$ into this expression, we obtain:
\begin{align}
    \langle C, \bar{T} \rangle - \epsilon H(\bar{T}) = \sum_{ij}  {C_{ij} \bar{T}_{ij}} + \bar{T}_{ij} \left( -(1-B^l_{ij})\Lambda^B_{ij} + \lambda^p_i + \lambda^q_j - c_{ij} - \epsilon \right),
\end{align}
from which, after developing the products and identifying dot products, we finally obtain the simplified expression:
\begin{align}
    \langle C, \bar{T} \rangle - \epsilon H(\bar{T}) = - \langle \bar{T}\odot (\mathds{1}_n\mathds{1}_m^\top-B^l), \Lambda^B \rangle + \langle\lambda^p, \bar{T} \mathds{1}_m \rangle +\langle\lambda^q, \bar{T}^\top \mathds{1}_n \rangle -\epsilon \langle \mathds{1}_n, \bar{T}\mathds{1}_m\rangle.
\end{align}
Finally, we may inject this result into the dual formulation from~\eqref{eq:dual} to obtain 
\begin{align}
    \mathcal{OT}_\epsilon^l(p, q)
    &= \max\limits_{\lambda^p, \lambda^q, \Lambda^B} \langle \lambda^p, p \rangle + \langle \lambda^q, q \rangle -\epsilon \langle \mathds{1}_n,  \bar{T}_{ij}(\bm{\Lambda})\mathds{1}_m\rangle\label{eq:lot_plugged_t}.
\end{align} 
To solve this optimization problem, we first define the objective of \eqref{eq:lot_plugged_t} as $\mathcal{L}_\lambda$.
\begin{align}
    \mathcal{L}_{\lambda}(\bm{\Lambda}) := \langle \lambda^p, p \rangle + \langle \lambda^q, q \rangle -\epsilon \sum_{ij} \bar{T}_{ij} (\bm{\Lambda})
\end{align}
We now seek to find a critical point for variable $\Lambda^B$ for fixed values of $\lambda_p$ and $\lambda_q$:
\begin{align}
\frac{\partial \mathcal{L}_\lambda}{\partial \Lambda^B_{ij}} = 
 {\exp\left(\frac{1}{\epsilon}\left(-(1-B^l_{ij})\Lambda^B_{ij} + \lambda^p_i + \lambda^q_j - C_{ij}\right)\right)}(1-B_{ij}^l) 
 \label{eq:diff}
\end{align}
We now notice that for any pairs of indices $(i, j)$ such that $B^l_{ij} = 0$, the partial derivative expressed in \eqref{eq:diff} is positive for all values of $\bm{\Lambda}$, but vanishes for $\Lambda^B_{ij} \rightarrow +\infty$ for any fixed values of $\lambda_p$ and $\lambda_q$. Then, we notice that for any pairs of indices $(i, j)$ such that $B^l_{ij} = 1$, \eqref{eq:lot_plugged_t} is constant with respect to $\Lambda^B_{ij}$, therefore, in that case we may pick $\Lambda^B_{ij} = 0$.

Plugging the optimal values for $\Lambda^B_{ij}$ in \eqref{eq:ot_plugged_t}, we obtain
\begin{align}
    \mathcal{OT}_\epsilon^l(p, q)
    &= \max\limits_{\lambda^p, \lambda^q} \langle \lambda^p, p \rangle + \langle \lambda^q, q \rangle + \langle R^\epsilon(\lambda^p, \lambda^q, C), B^l \rangle, \label{eq:new_opt}
\end{align}
where $[R^\epsilon(\lambda^p, \lambda^q, C)]_{ij} = \exp(\frac{1}{\epsilon}(\lambda^p_i + \lambda^q_j - c_{ij}))$. Defining $({\lambda^p}^*, {\lambda^q}^*)$ as the optimal solution for $\eqref{eq:new_opt}$, we now get the optimal transport plan $T^*$ for \eqref{eq:dual}:
\begin{align}
    T_{ij}^*({\lambda^p}^*, {\lambda^q}^*) &= \begin{cases}  \exp(\frac{1}{\epsilon}({\lambda^p_i}^* + {\lambda^q_j}^* - C_{ij})), & \text{if } B_{ij} = 1 \\ 0, & \text{otherwise}
    \end{cases} \nonumber \\
    &= \textrm{diag}(a)(K\odot B^l)\textrm{diag}(b) \nonumber
\end{align}
where 
\begin{align}
a_i = \exp\left(\frac{{\lambda^p_i}^*}{\epsilon}\right), \ b_j = \exp\left(\frac{{\lambda^q_j}^*}{\epsilon}\right), \ K_{ij} = \exp\left(-\frac{C_{ij}}{\epsilon}\right),
\end{align}
which completes the proof.
\end{proof}

Additionally, we note that $K\odot B^l$ can be the input for the Bregman projections described in Remark 4.8 from \citet{Peyre2018-pb}. In such case, the Sinkhorn iterations will converge to the solution of the optimization problem. 

\section{Cost calculation of labeled GWOT} \label{app:cost}
We present how to speed up the calculation of the cost in the case of labeled GWOT when the cost functions $d$ can be written as 

\costclass*

This includes a wide array of cost functions, including the $L^2$ distance $d(a, b) = \|a\|^2 + \|b\|^2 - 2\|a\|\|b\|$. For such functions, \citet{Peyre2016-jg} proposed an algorithm to speed up the cost calculations of GWOT. We first present this development, and then show how it can be adapted to the labeled GWOT problem to further improve the speed up.

Recall the cost of GWOT is written as 
\begin{align}
    \min_{T\in\mathcal{C}_{p,q}}{\sum_{i,j, k,l} \mathcal{D}_{ijkl} T_{ij} T_{kl}}, \label{eq:gwot_cost}
\end{align}
where $\mathcal{D}_{ijkl} = c(c_\mathcal{X}(x_i, x_k), c_\mathcal{Y}(y_j, y_l))$, and $T \in \mathbb{R}^{n \times m}$.
With tensor-matrix multiplication $\mathcal{D}\otimes T \in \mathbb{R}^{n \times m}$ defined on 4-dimensional tensor $\mathcal{D}\in \mathbb{R}^{n\times m\times n' \times m'}$ defined as
\(
    [\mathcal{D}\otimes T]_{ij} := \sum_{k,l} \mathcal{D}_{ijkl} T_{kl},
\)
the objective function of \eqref{eq:gwot_cost} can be written as
\(
    \langle \mathcal{D}\otimes T, T\rangle.
\)

\citet{Peyre2016-jg} showed that for the cost function $d$ in the form \eqref{eq:cost_class}, $\mathcal{D} \otimes T$ can be simplified as follows. Let  $M_{ij} = c_{\mathcal{X}}(x_i, x_j)$ and $\bar{M}_{ij} = c_{\mathcal{Y}}(y_i, y_j)$ denote the cost matrices on each modality. Let us define $c_{M, \bar{M}} = f_1(M)p\mathds{1}_m^\top + \mathds{1}_n q^{\top}f_2(\bar{M})^\top$, and notice that
\begin{align}
    \langle \mathcal{D}\otimes T, T\rangle = \langle c_{M, \bar{M}} - h_1(M)Th_2(\bar{M})^\top, T\rangle. \label{eq:kernel_cost}
\end{align}
Using this formulation, the cost $\mathcal{D}\otimes T$ may be calculated in time $O(n^2m + nm^2)$, instead of $O(n^2m^2)$. 

We now show that the calculation of $\mathcal{D} \otimes T$ in \eqref{eq:kernel_cost} can be further accelerated when $T \in \mathcal{C}_{p,q}^l$. Recall the notation for the indices of the source and target samples of a given label $a$ were $l_x\inv (a) = \{i \mid l^x_i = a\}$ and $l_y\inv (a) = \{j \mid l^y_j = a\}$, respectively. Here, we further denote for each label $k$, $n^a = |l_x\inv (a)|$ and $m^a = |l_y\inv (a)|$ as the number of samples with that particular label in modality $\mathcal{X}$ and $\mathcal{Y}$, respectively. 

Recall that for a matrix $A$, we denote as $[A]_{\{i_1, \ldots, i_n\}, \{j_1, \ldots, j_m\}}$ the submatrix of $A$ with $i_1, \ldots, i_n$-th rows and $j_1, \ldots, j_m$-th columns. Several submatrices are now introduced to calculate the optimal transport cost at the resolution of the labels. We denote as $T_a \in \mathbb{R}^{n^a \times m^a}$, the submatrix corresponding to the label-specific coupling:
\begin{align}
    T_k \coloneq& [T]_{l_x\inv (a), l_y\inv (a)}.
\end{align}
Then, for a pair of labels $(a_1, a_2)$, we define the submatrices $M^{a_1a_2} \in \mathbb{R}^{n^a_1 \times n^{a_2}}$, $\bar{M}^{a_1a_2} \in \mathbb{R}^{m^a_1 \times m^{a_2}}$ as:
\begin{align}
    M^{a_1a_2} \coloneq& [M]_{l_x\inv (a_1), l_x\inv (a_2)} \\
    \bar{M}^{a_1a_2} \coloneq& [\bar{M}]_{l_y\inv (a_1), l_y\inv (a_2)}. 
\end{align}

With the definition, consider calculating \eqref{eq:kernel_cost} for $T\in \mathcal{C}_{p,q}^l$. We only need the $(i, j)$-th entries of $\mathcal{D}\otimes T$ for indices $(i, j)$ such that  $l^x_i = l^y_j$, because any other entry would not contribute to the final cost. We calculate the $[\mathcal{D}\otimes T]_{l_x\inv (a), l_y\inv (a)}$ for each label $a$ by calculating each terms of \eqref{eq:kernel_cost} for the label.

\begin{align}
    [c_{M, \bar{M}}]_{ij} &= \sum_{i'=1}^n f_1(M_{ii'})p_{i'} + \sum_{j'=1}^m f_2(\bar{M}_{jj'})q_{j'} \nonumber \\
    [c_{M, \bar{M}}]_{ij} &= \sum_{a'=1}^L \left(\sum_{i' \in l_x\inv(a')}f_1(M_{ii'})p_{i'} + \sum_{j' \in l_y\inv(a')} f_2(\bar{M}_{jj'})q_{j'} \right)
\end{align}
Writing this in a matrix form gives
\begin{equation}
    [c_{M, \bar{M}}]_{ l_x\inv (a),  l_y\inv (a)} = \sum_{a'=1}^L f_1(M^{aa'})p_{a'} + \sum_{a'=1}^L f_2(\bar{M}^{aa'})q_{a'} 
\end{equation}
The second term in \eqref{eq:kernel_cost} can be written as
\begin{equation}
    [h_1(M)Th_2(\bar{M})^\top]_{ij} = \sum_{i'=1}^n \sum_{j'=1}^m h_1(M_{ii'})T_{i'j'}h_2(\bar{M}_{jj'}). 
\end{equation}
As $T_{i'j'} = 0$ when $l^x_i \neq l^y_j$,
\begin{equation}
    [h_1(M)Th_2(\bar{M})^\top]_{ij} = \sum_{a=1}^L \sum_{i' \in \\ l_x\inv (a)} \sum_{j' \in \\ l_y\inv (a)} h_1(M_{ii'})T_{i'j'}h_2(\bar{M}_{jj'}).
\end{equation}
Writing this in a matrix form gives:
\begin{equation}
    [h_1(M)Th_2(\bar{M})^\top]_{ l_x\inv (a),  l_y\inv (a)} = \sum_{a'=1}^L h_1(M^{aa'})T^{aa'}h_2(\bar{M}^{aa'})^\top,
\end{equation}
which shows that the cost calculation of each label $a$ involves the cost between the samples of label $a$ and the samples of all other labels. Combining two terms gives
\begin{equation}
    [\mathcal{D}\otimes T]_{ l_x\inv (a),  l_y\inv (a)} = \sum_{a'=1}^L f_1(M^{aa'})p_{a'} + \sum_{a'=1}^L f_1(\bar{M}^{aa'})q_{a'} - \sum_{a'=1}^L h_1(M^{aa'})T^{aa'}h_2(\bar{M}^{aa'})^\top 
\end{equation}
With balanced number of samples $n^a = \frac{n}{L}$ and $m^a = \frac{m}{L}$, calculating \eqref{eq:cost_per_label} takes
\begin{equation*}
    \frac{n^2}{L} + \frac{m^2}{L} + \frac{n^2m}{L^2} + \frac{nm^2}{L^2}
\end{equation*}
operations, which reduces to
\begin{equation*}
    O\left(\frac{n^2m + nm^2}{L^2}\right).
\end{equation*}

Calculating this for all labels gives $\mathcal{D}\otimes T$ with
\begin{equation*}
    O\left(\frac{n^2m + nm^2}{L}\right)
\end{equation*}
operations, providing $L$ times acceleration compared to the original result from~\citet{Peyre2016-jg}.

\section{Block-level Sinkhorn update for Labeled Optimal Transport} \label{app:alg}
Assuming samples are sorted by their labels, As $K_{ij} = 0$ when $l^x_i \neq l^y_j$, we can write the Sinkhorn update in \cref{alg:egwl} line 9 as 
\begin{align}
a_i &\leftarrow \frac{p_i}{\sum_{j=1}^m K_{ij}B_{ij}b_j} = \frac{p_i}{\sum_{j\in \{j|l^x_i = l^y_j\}} K_{ij}b_j} \nonumber \\
b_j &\leftarrow \frac{q_j}{\sum_{i=1}^n K_{ji}B_{ji}a_i} = \frac{q_j}{\sum_{i\in \{i|l^x_i = l^y_j\}} K_{ji}a_i} \nonumber
\end{align}
Let $[v]_{\{i_1, i_2, \ldots, i_n\}}$ denote the smaller vector $(v_{i_1}, \ldots, v_{i_n})$ consisted of the $i_1, \ldots, i_n$-th entries of a vector $v$. The updates can be written in the matrix form for the entries corresponding to a label $k$, where $\varoslash$ denotes the element-wise division.
\begin{align}
[a]_{l_x\inv (k)} &\leftarrow [p]_{l_x\inv (k)} \varoslash [K]_{l_x\inv (k), l_y\inv (k)}[b]_{l_y\inv (k)} \nonumber \\
[b]_{l_y\inv (k)} &\leftarrow [q]_{l_y\inv (k)} \varoslash ([K]_{l_x\inv (k), l_y\inv (k)})^\top [a]_{l_x\inv (k)} \nonumber
\end{align}
With the balanced number of samples per label $n^k = n / L$ and $m^k = m / L$, updating for the entries corresponding to a single label is $O(nm/L^2)$ and the overall update for all labels is $O(nm/L)$, providing $L$ times speedup compared to standard Sinkhorn update (for non-$l$-compatible $T$).

\section{Computational Complexity for Labeled COOT} \label{app:cootl}
Recall \cref{alg:ecootl}.
\setcounter{algorithm}{1}
  \begin{algorithm}[H]
   \caption{BCD algorithm for Labeled entropic COOT}
\begin{algorithmic}[1]
   \STATE {\bfseries Input:} $\mathcal{K}, \mathcal{K}', l, p, q, \epsilon^v, \epsilon^{s(1)},\ldots, \epsilon^{s(L)}$
   \STATE Initialize $T^{s(1)}$, \ldots, $T^{s(L)}$, $T^v$.
   \REPEAT
        \STATE $T^{v} \leftarrow \mathcal{T}_{\epsilon^v}(r, t, \sum_{a=1}^L\mathcal{K}'^a\otimes {T^s}^{(a)})$.
        \FOR{$a=1$ {\bfseries to} $L$}
            \STATE ${T^s}^{(a)} \leftarrow \mathcal{T}_{{\epsilon^s}^{(a)}}(p^a, q^a, \mathcal{K} \otimes T^v)$.
       \ENDFOR
   \UNTIL convergence
\end{algorithmic}
\end{algorithm}
Line 4 of Algorithm \ref{alg:ecootl} can be obtained from the $T^v$ update \eqref{eq:coot_update} in Algorithm 1 of \citet{Redko2020-bn}.
\begin{equation}
    T^v \leftarrow \mathcal{T}_{\epsilon^v}(r, t, \mathcal{K}' \otimes T^s) \label{eq:coot_update}
\end{equation}
As $T^s \in \mathcal{C}_{p,q}^l$ and $T^s_{ij} = 0$ for $(i, j) \notin \{i, j \mid l^x_i = l^y_j \}$,
\begin{equation}
    [\mathcal{K}' \otimes T^s]_{kl}= \sum_{i,j} \mathcal{K}'_{klij} T^s_{ij} 
    = \sum_{a=1}^L \sum_{i \in l_x\inv (a)} \sum_{j \in l_y\inv (a)} \mathcal{K}'_{klij} T_{ij}
    = \sum_{a=1}^L \mathcal{K}'^a \otimes {T^s}^{(a)}. \nonumber
\end{equation}

\subsection{Time Complexity}

Assuming the cost function $c$ to follow the form in \eqref{eq:cost_class}, the tensor multiplication in lines 4 and 6 of \cref{alg:ecootl} can be written as follows:
\begin{equation}
    \mathcal{K}'^a\otimes {T^s}^{(a)} = X^{a\top} p^a \mathds{1}_{d_2}^\top + \mathds{1}_{d_1} q^{a\top} Y^{a\top} \nonumber - h_1(X^{a\top}){T^s}^{(a)}h_2(Y^{a\top})^\top
\end{equation}
\begin{equation}
    \mathcal{K}^a\otimes {T^v} = X^a r \mathds{1}_{m/L}^\top + \mathds{1}_{n/L} t Y^{a\top} \nonumber- h_1(X^a){T^v}h_2(Y^a)^\top \nonumber
\end{equation}
Thus, line 4 takes
\begin{equation*}
    L \left( d_1\frac{n}{L}\frac{m}{L} + d_1\frac{m}{L}d_2 \right)
\end{equation*}
operations and line 6 takes
\begin{equation*}
    \frac{n}{L}d_1d_2 + \frac{n}{L}d_2\frac{m}{L}
\end{equation*}
operations for each of $L$ labels.

Overall, the time complexity of the cost calculation for the labeled COOT is
\begin{equation*}
    O\left(\frac{nm(d_1+d_2)}{L} + d_1d_2(n+m)\right).
\end{equation*}

However, COOT without sample labels would take
\begin{equation*}
    d_1nm + d_1md_2 \quad \text{and} \quad nd_1d_2 + nd_2m
\end{equation*}
operations, leading to
\begin{equation*}
    O(nm(d_1+d_2) + d_1d_2(n+m)).
\end{equation*}

Thus, labeled COOT provides $L$ times speedup compared to COOT when $n, m \gg d_1, d_2$.

\subsection{Space Complexity}
Space complexity of the above cost calculations is
\begin{equation*}
    O\left(\max\left(d_1\frac{n}{L} + \frac{n}{L}\frac{m}{L}, d_1\frac{m}{L} + d_2\frac{m}{L}, d_1\frac{n}{L} + d_1d_2, d_2\frac{n}{L} + d_2\frac{m}{L}\right)\right).
\end{equation*}

\section{Experimental details} \label{app:exp}
\begin{figure}[ht]
\centering
\includegraphics[width=400pt]{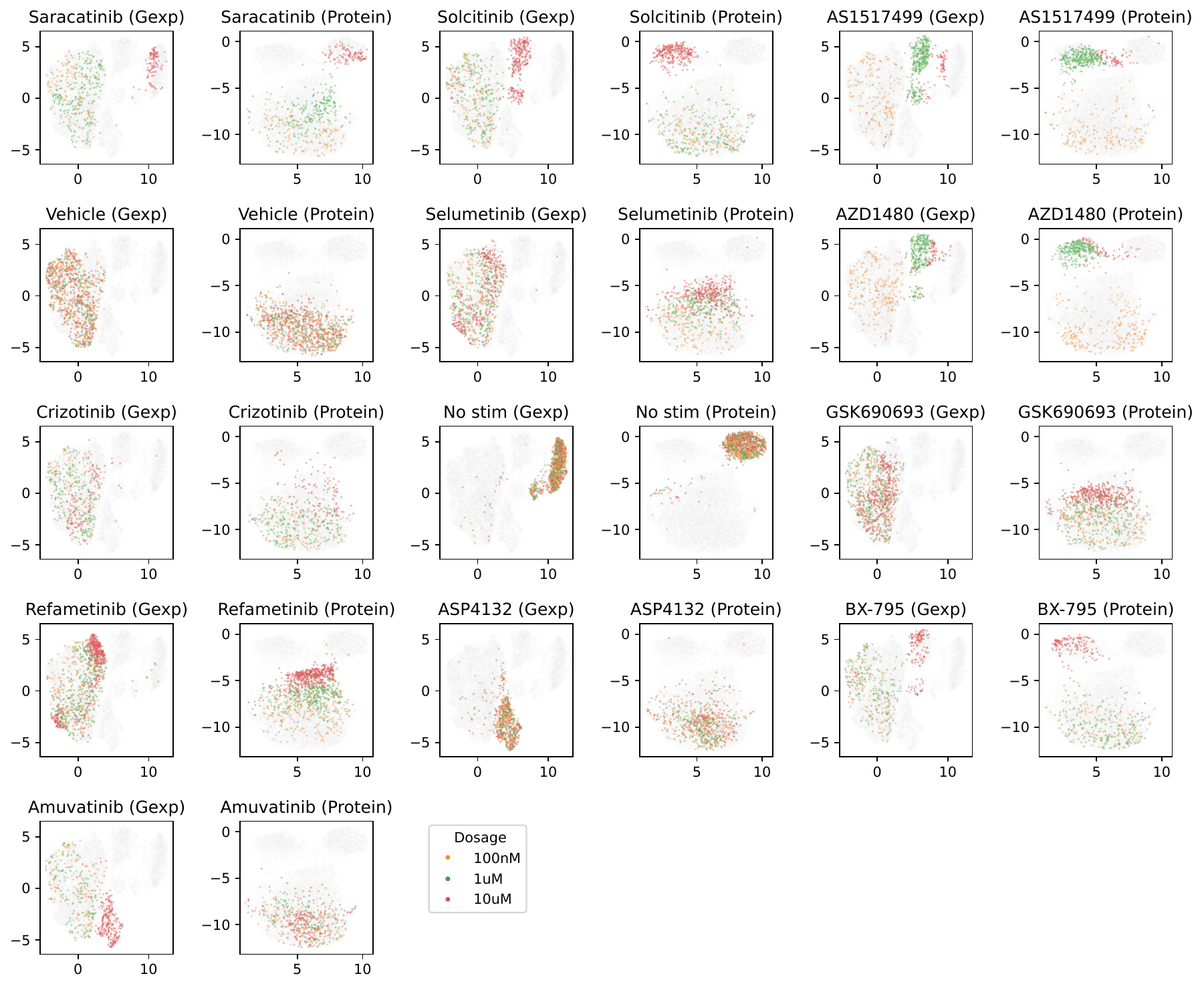}
\caption{UMAPs of each modality for the cells treated with the selected 13 kinase inhibitors.}
\label{input_umap}
\end{figure}

\paragraph{Multi-omics single-cell data processing} 
Gene expression (RNA) levels from the kinase inhibitor screening data were library size normalized and log transformed as defaults, and the top 2000 highest variable genes were retained using the `\verb+scanpy+` Python library\citep{Wolf2018-gl}. Protein features were manually selected by visual inspection of changes across control perturbations, retaining 123 of 277 measured proteins. Protein modality counts were transformed using centered log ratios (CLRs), as explained in~\citet{Chung2021-wy}. The first 50 PCs were obtained at this step with the `\verb+sc.tl.pca+` command from \citet{Wolf2018-gl}. 11 kinase inhibitors with large effects on the PCA space across dosages were selected (Figure \ref{input_umap}), along with vehicle treatment and non-stimulated T cells as the negative controls.

\paragraph{Sample matching matrix for DAVAEs}
Dosage match scores and feature matching task requires sample matching matrices. For DAVAEs (see \cref{app:davae}), sample matching matrices were calculated \textit{post-hoc} by obtaining coupling from the latent representations. Coupling matrices were obtained by the `\verb+sc.neighbors._connectivity.gauss+` command from \citet{Wolf2018-gl} with $k$ across $\{1, 5, 10, 20, 30\}$ and the highest dosage match score was used.

\paragraph{Model structure and training} 
All OT-based methods were cost-normalized so that the maximum cost is $1$ and allowed for a maximum 2000 iterations both for inner and outer iterations, if applicable.
MLPs had 2 hidden layers with 123 dimensions with a batch normalization and ReLU activation layer, followed by a dense linear output layer. Mean squared error is minimized with an Adam optimizer with a learning rate of $10^{-3}$ and early stopping when the validation loss does not decrease for 45 epochs with max 2000 epochs. 

\paragraph{Feature matching} 
We obtained the feature coupling by plugging $T^s$ in \eqref{eq:ecoot}, resulting in the following EOT problem between features:
\begin{equation}
    \min_{T^v \in \mathcal{C}_{r,t} }\sum_{i, j , k, l} c(x_{ik}, y_{jl}) T^v_{kl}T^s_{ij} - \epsilon^v H(T_v). \nonumber
\end{equation}
We used the 23 proteins/RNAs that were measured in both modalities: CD70, CD52, CD7, TIGIT, CD69, CTLA4, LAG3, CD27, Fas, BTLA, ITGB7, CD83, CXCR4, CD55, CD38, CD9, CD109, CD84, FOXP3, CTLA4, IRF4, GATA3, and FOXP3.

\section{Results on Synthetic Data}
\label{app:simulation}

\subsection{Data Generating Process}
To simulate single-cell multi-modal perturbation responses, we assume a latent, lower-dimensional variation creates multi-modal views with higher-dimensional features. We further assume a "program-level" perturbation induced by chemical perturbations, where each perturbation shifts one of the latent dimensions. 

We begin by defining the latent variables. Let $Z$ represent the latent variation, sampled from a normal distribution:
\[
Z \sim \mathcal{N}(0, 0.1),
\]
such that $Z \in \mathds{R}^{n \times d}$, where $n$ is the number of cells and $d$ is the dimensionality of the latent space.

Each modality is associated with a transformation matrix $A$. For modality $X$, $A_X \in \mathds{R}^{d \times p_X}$, where $p_X$ is the number of features in the modality:
\[
A_X \sim \mathcal{N}(0, 1).
\]
Similarly, $A_Y$ corresponds to modality $Y$, with $A_Y \in \mathds{R}^{d \times p_Y}$ and $p_Y$ features:
\[
A_Y \sim \mathcal{N}(0, 1).
\]

Bias vectors $b_X$ and $b_Y$ are defined for each modality, with $b_X \in \mathds{R}^{p_X}$ and $b_Y \in \mathds{R}^{p_Y}$:
\[
\begin{aligned}
b_X &\sim \mathcal{N}(0, 1), \\
b_Y &\sim \mathcal{N}(0, 1).
\end{aligned}
\]

Scaling factors $s_X$ and $s_Y$ control the variability of features within each modality. These factors are drawn from a Gamma distribution, where $s_X \in \mathds{R}^{p_X}$ and $s_Y \in \mathds{R}^{p_Y}$:
\[
\begin{aligned}
s_X &\sim \Gamma(1, 1), \\
s_Y &\sim \Gamma(1, 1).
\end{aligned}
\]

To account for variability in the latent space, we introduce modality-specific perturbations $\zeta_X$ and $\zeta_Y$. Their means $\mu_X$ and $\mu_Y$ are drawn from a standard normal distribution, while their log-variances $\log(\sigma_X)$ and $\log(\sigma_Y)$ are drawn from a normal distribution with mean $-3$ and variance $0$:
\[
\begin{aligned}
\mu_X &\sim \mathcal{N}(0, 1), &\quad \log(\sigma_X) &\sim \mathcal{N}(-3, 0), \\
\mu_Y &\sim \mathcal{N}(0, 1), &\quad \log(\sigma_Y) &\sim \mathcal{N}(-3, 0).
\end{aligned}
\]
The perturbations are sampled using these parameters:
\[
\begin{aligned}
\zeta_X &\sim \mathcal{N}(\mu_X, \sigma_X), \\
\zeta_Y &\sim \mathcal{N}(\mu_Y, \sigma_Y).
\end{aligned}
\]

Using these components, the observed modalities $X$ and $Y$ are generated as functions of $Z$:
\[
\begin{aligned}
X &= ((Z + \zeta_X)A_X + b_X) s_X, \\
Y &= ((Z + \zeta_Y)A_Y + b_Y) s_Y.
\end{aligned}
\]

Next, we incorporate perturbations. We define $L = 10$ perturbations, each mapped to a target latent dimension $t(\mathrm{I})$ based on the index modulo the number of dimensions, $d$. Let:
\[
t(\mathrm{I}) = ((\mathrm{I} - 1) \bmod d) + 1,
\]
where $t(\mathrm{I}) \in \{1, 2, \dots, d\}$ ensures a cyclic mapping of perturbation indices to dimensions.

The effect size $e(\mathrm{I})$ for each perturbation is drawn from a Gamma distribution with a minimum value of 3:
\[
e(\mathrm{I}) \sim \max(3, \Gamma(1, 1)).
\]

To model variability in the perturbation response \citep{Goodrich2021-fs}, each cell has a penetrance $q_i$ sampled from a Beta distribution:
\[
q_i \sim \mathrm{Beta}(1, 10).
\]
The perturbed latent space $z_i^\mathrm{I}$ for cell $i$ under perturbation $\mathrm{I}$ is defined by adjusting the targeted dimension:
\[
z_i^\mathrm{I} = (z_{i1}, \dots, z_{it(\mathrm{I})} + e(\mathrm{I}) \times q_i, \dots, z_{id}).
\]
The perturbed observations for modalities $X$ and $Y$ are then computed as:
\[
\begin{aligned}
x_i^\mathrm{I} &= f_X(z_i^\mathrm{I}), \\
y_i^\mathrm{I} &= f_Y(z_i^\mathrm{I}).
\end{aligned}
\]

We simulate one untreated condition and nine independent perturbations, each with 50 cells. The simulated dataset has  5-dimensional latent variation, viewed in
two modalities each with 50 and 200 features.

\subsection{Results}
We present the results for the sample matching task in \cref{tbl:synthetic_metrics_matching}, and those for the prediction task in \cref{tbl:synthetic_metrics_prediction}.

\begin{table}[p]
\vspace{-0.15in}
\caption{Evaluation metrics of OT and GW approaches for sample matching task on the synthetic dataset.}
\label{tbl:synthetic_metrics_matching}
\begin{center}
\begin{tabular}{rlccc}
\toprule
\multicolumn{2}{c}{Method} & \makecell{Bary\\FOSTTM($\downarrow$)} & \makecell{Mean\\rank} \\
\midrule
\multicolumn{2}{c}{Perfect}& 0 &  -\\
\multicolumn{2}{c}{Uniform per label}& 0.256 &  - \\ 
\midrule
EOT     & no label & 0.375 & 9\\ 
        & per label     & 0.238 & 7\\ 
ECOOT   & no label & 0.479  & 10 \\ 
        & per label     & 0.207  & 4 \\ 
        & labeled       & 0.167 & 3 \\
EGWOT   & no label & 0.212 & 5 \\ 
        & per label     & \underline{0.118} & 2 \\
        & labeled       & \textbf{0.045} & 1 \\
DAVAE   & no label & 0.226 & 6\\ 
        & labeled      & 0.32 & 8 \\  
\bottomrule
\end{tabular}
\end{center}
\vspace{-0.15in}
\end{table}

\begin{table*}[p]
\caption{Evaluation metrics of OT and GW approaches for prediction task on the synthetic dataset.}
\label{tbl:synthetic_metrics_prediction}
\begin{center}
\begin{tabular}{rlcccccc}
\toprule
\multicolumn{2}{c}{Method} & $R_v$ ($\uparrow$) & $\rho_v$ ($\uparrow$) & $R_s$ ($\uparrow$) & $\rho_s$ ($\uparrow$) & MSE ($\downarrow$) & \makecell{Mean\\rank} \\
\midrule
\multicolumn{2}{c}{Perfect}& 0.533 & 0.55 & 0.634 & 0.621 & 0.18 & - \\ 
\multicolumn{2}{c}{Uniform per label}& 0.257 & 0.329 & 0.354 & 0.352 & 0.294 & - \\ 
\midrule
EOT     & no label & 0.122 & 0.095 & 0.087 & 0.083 & 0.568 & 7 \\ 
        & per label     & 0.244 & 0.295 & 0.34 & 0.33 & 0.316 & 5\\ 
ECOOT   & no label & 0.063 & -0.067 & -0.096 & -0.087 & 0.647 & 9.2\\ 
        & per label     & 0.246 & 0.292 & 0.335 & 0.321 & 0.325 & 5.6\\ 
        & labeled       & \underline{0.353} & 0.354 & 0.413 & 0.404 & \underline{0.287} & 3 \\ 
EGWOT   & no label & 0.172 & 0.339 & 0.402 & 0.393 & 0.282 & 4.2\\ 
        & per label     & 0.296 & \underline{0.407} & \underline{0.455} & \underline{0.444} & 0.268 & 2.2\\ 
        & labeled       & \textbf{0.375} & \textbf{0.484} & \textbf{0.566} & \textbf{0.554} & \textbf{0.22} & 1 \\ 
DAVAE   & no label & 0.04 & 0.024 & 0.032 & 0.038 & 1.874 & 8.4\\ 
        & labeled       & -0.03 & 0.002 & 0.0 & 0.003 & 1.895 & 9.4\\ 
\bottomrule
\end{tabular}
\end{center}
\end{table*}

\section{Labeled domain-adversarial VAE} \label{app:davae}
We have modified MultiVI \citep{Ashuach2023-xu} to account for labels when calculating adversarial loss. Here, the adversarial classifier accepts the label to promote matching modality per label with the following loss.
\begin{equation}
 \mathop{\min}\limits_{\theta, \phi} \max\limits_{ \zeta} \mathcal{L}
= \mathop{\min}\limits_{\theta, \phi} \max\limits_{ \zeta}\mathcal{L}^{\text{ELBO}}_{\theta, \phi}(X, Y) + \lambda \mathcal{L}^{Adv}_{\zeta, \phi}(X, Y, l)  \label{eq:vae}
\end{equation}
Loss terms of \eqref{eq:vae} are as follows.
\begin{align}
\mathcal{L}^{\text{ELBO}}_{\theta, \phi}(X, Y) =& -(\text{ELBO}_{\theta_X, \phi_X}(X) + \text{ELBO}_{\theta_Y, \phi_Y}(Y)) \nonumber \\
\mathcal{L}^{Adv}_{\zeta, \phi}(X, Y, l) =& \sum_{x_i \in X} \frac{1}{|X|}\text{CELoss}(f_\zeta(q_\phi(x_i), l^x_i), \langle 0, 1 \rangle) \nonumber\\
&+ \sum_{y_j \in Y} \frac{1}{|Y|}\text{CELoss}(f_\zeta(q_\phi(y_j), l^y_j), \langle 1, 0 \rangle)\nonumber \\
\text{CELoss}(\vec{x}, \vec{t} ) =& -\sum_{i=\{0,1\}}t_i \log\left( \frac{e^{x_i}}{\sum_{k=\{0,1\}} e^{x_k}} \right) \nonumber 
\end{align}
For the DAVAE without label adaptation, $f_\zeta$ does not admit label $l$ of samples, and the rest of the structure is the same.

\paragraph{Experimental details}
DAVAE and labeled DAVAE used normal likelihood for $X$ and $Y$:
\begin{equation}
x_i \mid z_i \sim \mathcal{N}(\mu_i, \sigma^2)
\end{equation}
where $x_i$, $z_i$, $\mu_i$, $\sigma^2$ are length $d$ vectors with $d$ number of features. The log-transformed standard deviations ($\log(\sigma_j)$ for feature $j \in \{1, \cdots, d\}$) were learned as parameters. VAEs for both modalities had 2 hidden layers for both the encoder and decoder and 50 latent dimensions. Each hidden layer of the VAEs for the protein and RNA modalities had 128 and 256 dimensions, respectively. The adversarial classifier had 3 hidden layers with 32 dimensions. The model was trained with early stopping when validation loss did not decrease for 50 epochs and a maximum of 2000 epochs. The Adam optimizer for $\theta, \phi$ used a learning rate of $10^{-4}$ and the Adam optimizer for $\zeta$ used a learning rate $10^{-3}$. The remaining parameters were the same as the original MultiVI defaults.

\section{Extended Results on Real Data} \label{app:result}

\begin{figure}[t] 
\centering
\includegraphics[height=0.95\textheight]{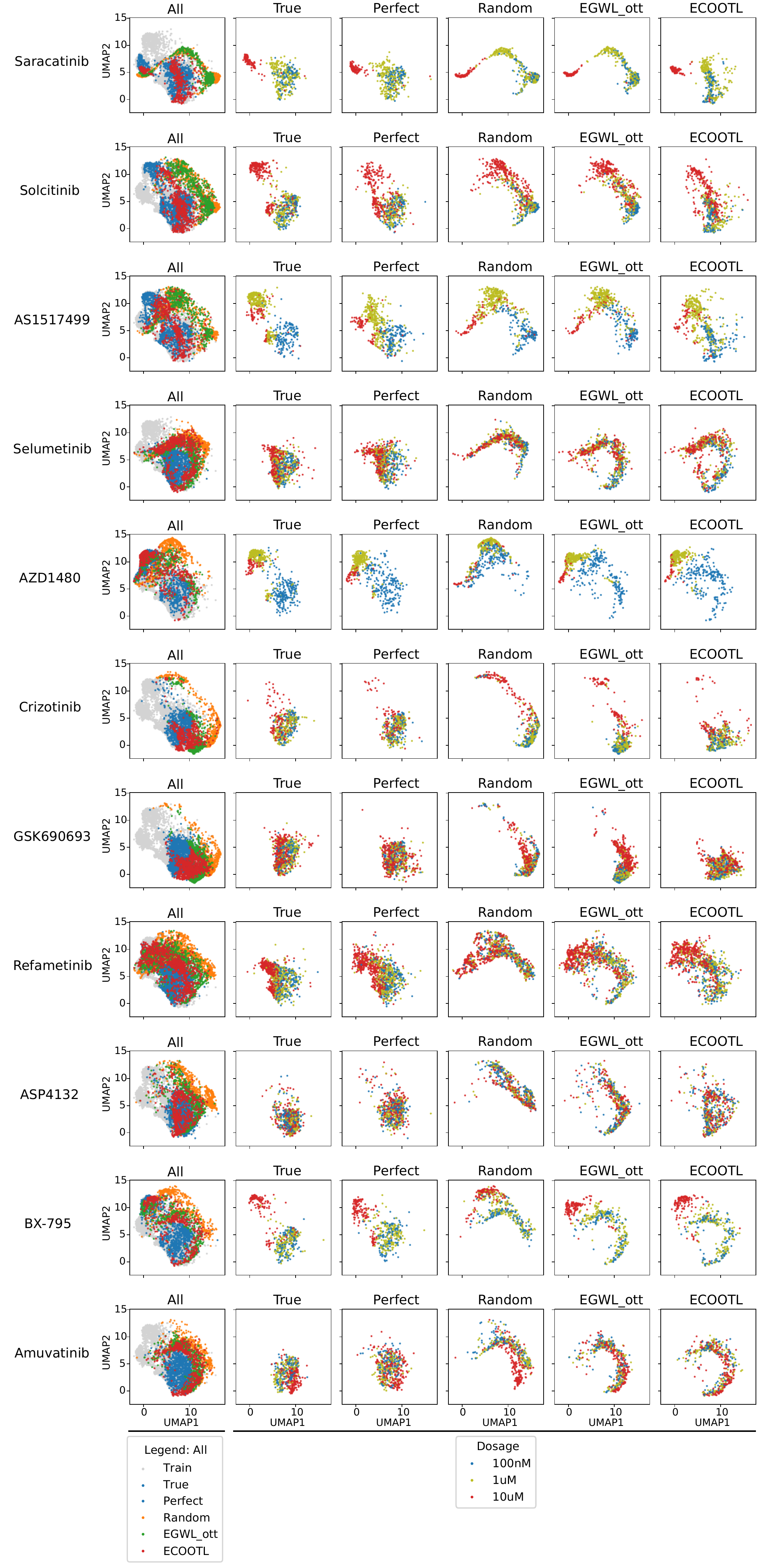}
\caption{UMAPs of predicted cells}
\label{fig:pred_umap}
\end{figure}

\paragraph{UMAPs of predicted cells}
We show the UMAP embeddings of the true and predicted cell profiles for the best-performing methods in Figure \ref{fig:pred_umap}.

\begin{figure}[t] 
\centering
\includegraphics[width=\columnwidth]{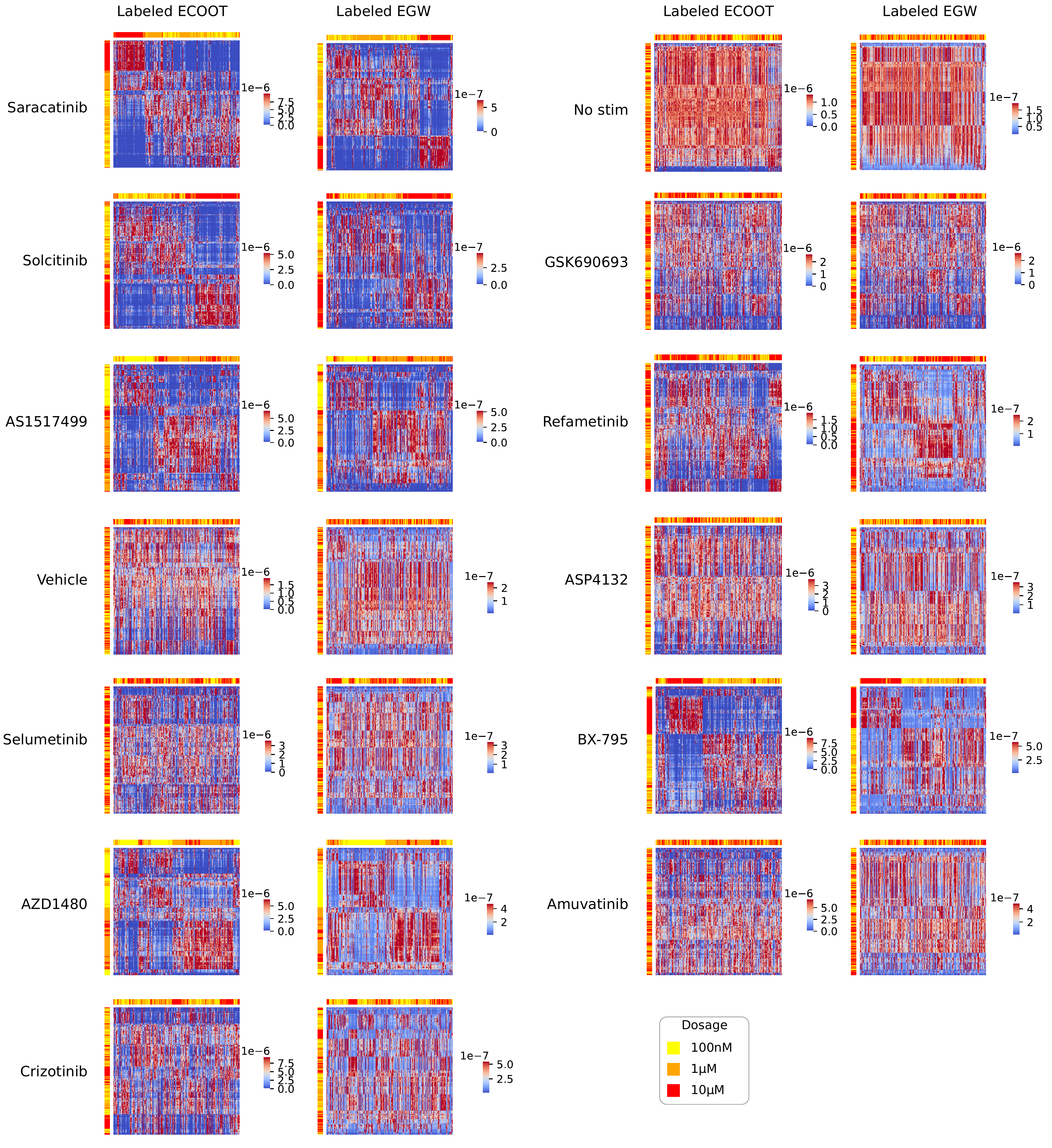}
\caption{Mean couplings of ECOOTL and EGWL for 5 cross-validation folds. Rows are hierarchically clustered and columns are reordered by the same row order. Rows and columns are labeled by the dosage.}
\label{fig:match}
\end{figure}

\paragraph{Match matrices}
We show the mean couplings of ECOOTL and EGWL for 5 cross-validation folds in Figure \ref{fig:match}. Whereas the negative controls (vehicle and non-activation) showed no clear separation of samples by the treatment dosages, we see the clustering of cells with the same dosages of the drugs.

\paragraph{Full evaluation metrics}
We show the full evaluation metrics with standard deviations across 5 cross validation folds for prediction and matching tasks in \cref{tbl:metrics_matching_std,tbl:metrics_prediction_std}.

\begin{table}[h]
\caption{Evaluation metrics of OT and GW approaches for sample matching task with standard deviations.}
\label{tbl:metrics_matching_std}
\begin{center}

\begin{tabular}{rlccc}
\toprule
\multicolumn{2}{c}{Method} & \makecell{Bary\\FOSTTM($\downarrow$)} & \makecell{Dosage\\match ($\uparrow$)}& \makecell{Mean\\rank} \\
\midrule
\multicolumn{2}{c}{Perfect}          & $0 \pm 0 $    & $1 \pm 0$ & -\\
\multicolumn{2}{c}{By dosage}        & $0.239 \pm 0.018$   & $1 \pm 0$ & -\\
\multicolumn{2}{c}{Uniform per label}& $0.298 \pm 0.020$   & $0.357 \pm 0.003$ & -\\
\midrule
EOT     & no label & $0.428 \pm 0.013$ & $0.040 \pm 0.003$& 10 \\
        & per label     & $0.336 \pm 0.019$ & $0.346 \pm 0.006$ & 5.5 \\
ECOOT   & no label & $0.414 \pm 0.047$ & $0.049 \pm 0.005$ & 9 \\
        & per label     & $0.359 \pm 0.019$ & $0.380 \pm 0.019$ & 5.5 \\
        & labeled       & $0.270 \pm 0.039$ & $\textbf{0.456} \pm 0.039$ &  2 \\
EGWOT   & no label & $0.373 \pm 0.017$ & $0.068 \pm 0.029$ & 8 \\
        & per label     & $0.332 \pm 0.021$ & $0.381 \pm 0.018$ & 4 \\
        & labeled       & $0.283 \pm 0.027$ & $\underline{0.452} \pm 0.006$ & 3 \\
DAVAE   & no label & $\textbf{0.231} \pm 0.056$ & $0.206 \pm 0.022$ & 3.5    \\
        & labeled       & $\underline{0.242} \pm 0.029$ & $0.205 \pm 0.021$ & 4.5  \\
\bottomrule
\end{tabular}
\end{center}
\vspace{-0.15in}
\end{table}

\begin{table}
\caption{Evaluation metrics of OT and GW approaches for prediction task with standard deviations.}
\label{tbl:metrics_prediction_std}
\begin{center}
\resizebox{\columnwidth}{!}{%
\begin{tabular}{rlcccccc}
\toprule
\multicolumn{2}{c}{Method} & $R_v$ ($\uparrow$) & $\rho_v$ ($\uparrow$) & $R_s$ ($\uparrow$) & $\rho_s$ ($\uparrow$) & MSE ($\downarrow$) \\
\midrule
\multicolumn{2}{c}{Perfect} & $0.107 \pm 0.015$ & $0.118 \pm 0.015$ & $0.163 \pm 0.016$ & $0.149 \pm 0.013$ & $0.258 \pm 0.013$ \\
\multicolumn{2}{c}{By dosage}& 0$.0812 \pm 0.016$ & 0$.0448 \pm 0.018$ & 0$.0903 \pm 0.017$ & 0$.0863 \pm 0.015$ & $0.264 \pm 0.014$ \\
\multicolumn{2}{c}{Uniform per label}& 0$.0794 \pm 0.018$ & 0$.0403 \pm 0.020$ & 0$.0761 \pm 0.013$ & 0$.0781 \pm 0.012$ & $0.264 \pm 0.014$ \\
\midrule
EOT     & no label & 0$.0482 \pm 0.005$ & $0.007 \pm 0.008$ & 0$.0068 \pm 0.014$ & 0$.0063 \pm 0.016$ & $0.287 \pm 0.010$ \\
        & per label     & 0$.0544 \pm 0.008$ & 0$.0239 \pm 0.010$ & 0$.0345 \pm 0.016$ & 0$.0307 \pm 0.016$ & $0.283 \pm 0.010$ & \\
ECOOT   & no label & $0.053 \pm 0.039$ & 0$.0207 \pm 0.039$ & 0$.0395 \pm 0.013$ & 0$.0408 \pm 0.015$ & $0.282 \pm 0.031$ \\
        & per label     & $0.0806 \pm 0.036$ & $0.0443 \pm 0.022$ & $0.0776 \pm 0.011$ & $0.0815 \pm 0.011$ & $0.265 \pm 0.013$ & \\
        & labeled       & $\textbf{0.0852} \pm 0.022$ & $\textbf{0.0523} \pm 0.022$ & $\textbf{0.0854} \pm 0.029$ & $\underline{0.0778} \pm 0.024$ & $\underline{0.265} \pm 0.017$ & \\
EGWOT   & no label & 0$.0631 \pm 0.026$ & 0$.0227 \pm 0.026$ & 0$.0302 \pm 0.018$ & $0.034 \pm 0.018$ & $0.282 \pm 0.020$ & \\
        & per label     & 0$.0785 \pm 0.016$ & $\underline{0.0449} \pm 0.019$ & 0$.0737 \pm 0.013$ & 0$.0737 \pm 0.014$ & $0.265 \pm 0.013$ & \\
        & labeled       & $\underline{0.0836} \pm 0.017$ & $0.044 \pm 0.019$ & $\textbf{0.0854} \pm 0.013$ & $\textbf{0.0825} \pm 0.010$ & $\textbf{0.264} \pm 0.014$ & \\
DAVAE   & no label & 0$.0342 \pm 0.015$ & -0$.0069 \pm 0.014$ & 0$.0006 \pm 0.006$ & -0$.0001 \pm 0.005$ &$ 0.33 \pm 0.013$\\
        & labeled       & 0$.0182 \pm 0.014$ & -0$.0079 \pm 0.015$ & -0$.0016 \pm 0.004$ & -0$.0014 \pm 0.004$ & $0.332 \pm 0.009$ \\
\bottomrule
\end{tabular}}
\end{center}
\end{table}

\end{document}